\documentclass[12pt,tightenlines,eqsecnum,floats,shownopacs,nofootinbib,amsmath,amssymb,aps,prd]{revtex4}

\usepackage{color}
\usepackage{graphicx}
\usepackage{amsmath,amssymb}
\newtheorem{theorem}{Theorem}

\usepackage[colorinlistoftodos]{todonotes}
\usepackage{epstopdf}

\def\be{\nopagebreak[3]\begin{equation}}
\def\ee{\end{equation}}
\def\ba{\nopagebreak[3]\begin{eqnarray}}
\def\ea{\end{eqnarray}}

\def\f{\frac}

\def\t{\tilde}
\def\h{\hat}
\def\sint{\textstyle{\int}}
\def\dd{{\rm d}}

\def\vx{\vec{x}}
\def\vp{\varphi}
\def\vk{\vec{k}}

\def\S{\mathcal{S}}

\def\ps{\Gamma}
\def\pscan{\Gamma_{{\rm Can}}}
\def\pscov{\Gamma_{{\rm Cov}}}
\def\psext{\Gamma_{{\rm Ext}}}
\def\H{\mathcal{H}}
\def\HJ{\mathcal{H}_{J}}
\def\He{\mathcal{H}_{\{e_{k}\}}}
\def\Hext{\mathcal{H}_{\rm Ext}}

\def\R{\mathcal{R}}

\def\Acan{\mathcal{A}_{\rm Can}}
\def\Acov{\mathcal{A}_{\rm Cov}}
\def\O{\mathcal{O}}
\def\J{\mathfrak{J}}
\def\g{\gamma}

\def\go{\mathring{g}}

\def\phio{\mathring{\phi}}
\def\varphio{\mathring{\varphi}}
\def\pio{\mathring{\pi}}
\def\Boxo{\mathring{\Box}}
\def\Jo{\mathring{J}}
\def\calJo{\mathring{\J}}
\def\pscano{\mathring{\Gamma}_{\rm Can}}
\def\pscovo{\mathring{\Gamma}_{\rm Cov}}
\def\Acano{\mathring{\mathcal{A}}_{\rm can}}
\def\Acovo{\mathring{\mathcal{A}}_{\rm cov}}
\def\Lambdao{\mathring{\Lambda}}
\def\Uo{\mathring{U}}

\def\gammao{\mathring{\gamma}}
\def\eo{\mathring{e}}
\def\Eo{\mathring{E}}

\begin{document}

\title{Unitarity and ultraviolet regularity in cosmology}

\author{Ivan Agullo$^{1}$}
\email{agullo@lsu.edu}
\author{Abhay Ashtekar$^{2}$}
\email{ashtekar@gravity.psu.edu}
\affiliation{$^{1}$ Department of Physics and Astronomy, The Louisiana State University, Baton Rouge, LA 70803}
\affiliation{$^{2}$ Institute for
Gravitation and the Cosmos \& Physics Department, Penn State, University Park, PA 16802, U.S.A.}

\begin{abstract}

Quantum field theory in curved space-times is a well developed area in mathematical physics which has had important phenomenological applications to the very early universe. However, it is not commonly  appreciated that on time dependent space-times ---including the simplest cosmological models--- dynamics of quantum fields is not unitary in the standard sense. This issue is first explained with an explicit example and it is then shown that a generalized notion of unitarity does hold. The generalized notion allows one to correctly pass to the Schr\"odinger picture starting from the Heisenberg picture used in the textbook treatments. Finally, we indicate how these considerations can be extended from simple cosmological models to general globally hyperbolic space-times.

\end{abstract}


\maketitle

\section{Introduction}
\label{s1}
Quantum field theory on cosmological space-times has been studied extensively both at the mathematical level, and in the context of the phenomenology of the early universe (see e.g. \cite{hw,parker-book,
ll-book,sd-book,vm-book,sw-book,gr-book,wald-book}). In the textbook formulation one generally works with \emph{space-time} field operators in the \emph{Heisenberg picture}. However, one can also recast the theory in a canonical framework and ask if the field operators and their conjugate momenta evolve via unitary transformations in the standard sense.   Unitary implementation of dynamics is essential, in particular, to pass to the Schr\"odinger picture. Rather surprisingly,  \emph{the answer is in the negative even for linear fields}.%
\footnote { This is an ultraviolet problem in the full quantum field theory. If one just restricts oneself to a finite number of modes one does, of course, have unitarity.}
Already in the late 1990s, this issue was discussed in a general context by Helfer \cite{ah}, and for evolution between arbitrary  Cauchy surfaces in flat space-time by Torre and Varadarajan \cite{torre-varadarajan}. It was made explicit in the cosmological context in a recent series of interesting papers by Corichi, Cortez, Mena-Marugan, Vehlinho and others, first in the Gowdy models \cite{corichietal}, then in the de Sitter space-time \cite{ds} and finally in Friedmann, Lema\^{i}tre, Robertson, Walker (FLRW) models \cite{cortezetal}. 

The purpose of this paper is to first explain the tension between unitarity and the textbook description of quantum fields on FLRW space-times, and then propose a resolution through a natural generalization of the standard notion of unitarity \emph{which does hold.} Some aspects of the necessity of a more general view of dynamics are present already in the classical Hamiltonian theory on cosmological backgrounds and in simpler quantum systems with time dependent Hamiltonians. However, we will find that there are also some important features which are special to \emph{quantum fields} propagating on \emph{time dependent space-times}. 

Let us begin by spelling out the conceptual tension in the simplest  setting: a scalar field $\phi$ satisfying $\Box \phi\, -\,  m^{2}\phi  =0$ in a spatially flat FLRW space-time. (The case $m=0$  is of direct interest to the analysis of tensor modes in the cosmological perturbation theory.) To emphasize the fact that the issue under consideration has its origin \emph{only} in the ultraviolet behavior of quantum fields ---and not infrared--- let us further suppose that the spatial topology is that of a 3-torus, $\mathbb{T}^{3}$, rather than $\mathbb{R}^{3}$.  

In the standard textbook treatment, one introduces the conformal time $\eta$ on space-time $\mathbb{M}$, so that the metric assumes the form 
\be ds^{2} = a^{2}(\eta)\, (-\dd \eta^{2} + \dd \vec{x}^{2})\, .\ee
One then expands the field operators $\h\phi(\vx,\eta)$ in terms of spatial Fourier modes
\be \label{fo}
\h\phi(\vx,\eta) = \frac{1}{V_{0}}\, \sum_{\vk}\, [e_{k}(\eta) \h{A}_{\vk}\, + \, e^{\star}_{k}(\eta) \h{A}^{\dag}_{-\vk} ]\,\, e^{i\vk\cdot\vx} \, ,\ee
with $k = |\vec{k}|$, and $V_{0}$, the volume of the 3-torus in the co-moving coordinates $\vx$. Here, the `positive frequency' basis $e_{k}(\eta)$ consists of functions that satisfy the equation of motion 
\be \label{e1}
e^{\prime\prime}_{k} (\eta)+ (2a^{\prime}/a)(\eta)\,\, e^{\prime}_{k} (\eta) + (k^{2} + m^{2} a^{2}(\eta)) \, e_{k}(\eta) =0\, ,\ee
and the normalization condition 
\be \label{e2}
 a^{2}(\eta)\, [ e_{k}e^{\star\prime}_{k} - e^{\prime}_{k} e^{\star}_{k}] = i\, , \ee
with prime denoting derivative with respect to $\eta$, and $\h{A}_{\vk}$ and $\h{A}^{\dag}_{\vk}$ are the annihilation and creation operators satisfying the standard commutation relations
\be [\h{A}_{\vk_{1}},\, \h{A}^{\dag}_{\vk_{2}}]\, = \, \hbar V_{0}\,\delta_{\vk_{1},\vk_{2}}\, . \ee
One defines $|0\rangle$ as the state annihilated by all $A_{\vk}$ and 
generates the Hilbert space of quantum states by repeatedly operating on it by creation operators. We will denote 
this space by $\He$ to emphasize the fact that this construction depends on the choice of a `positive frequency' basis $\{e_{k}\}$. Finiteness of the expectation values of the stress-energy tensor $\hat{T}_{ab}(\vx,\eta)$ on (a dense sub-space of) $\He$ is guaranteed if the basis $e_{k}(\eta)$ is adiabatic of order 4 or higher \cite{parker66,parker69,parker-fulling74,sf-book}.%
\footnote{For a succinct summary of the definition and properties of adiabatic states, see section VI of \cite{aan2}. In this paper, we  use the notions and structures spelled out there.}
From a physical perspective, then, it is natural to assume adiabiticity of order 4, although for most of our results adiabiticity only of order 2 will suffice. Of course, there is no preferred choice of the required basis $\{e_{k}(\eta)\}$, and hence of a vacuum state $|0\rangle$. Therefore notions such as `particle number' have an intrinsic ambiguity. However, one can unambiguously calculate expectation values of physical observables, such as the power spectrum and energy density, on (a dense subspace of) states in $\He$. Consequently, this textbook theory is deemed to be fully adequate to investigate the physics of the early universe.

But one can also examine this theory from the canonical perspective.
Using the expression (\ref{fo}) of the space-time field operator, we can introduce a pair of canonically conjugate operators at any \emph{fixed} time $\eta$:
\be \label{phipi}
\h\vp (\vx, \eta) := \h\phi(\vx, \eta) \quad {\rm and} \quad
\h\pi(\vx, \eta) := a^{2}(\eta)\,\, [{\partial \h\phi}/{\partial \eta}] (\vx,\eta)\, .  \ee
From their definition it is obvious how these operators evolve in conformal time $\eta$. Is this evolution unitary? That is, does there exist a 2-parameter family of unitary operators $U_{\eta_{2}, \eta_{1}}$ on $\He$ that implement this dynamics in the standard sense
\be \label{unitary}
\h\vp (\vx,\eta_{2}) = U_{\eta_{2},\eta_{1}}^{-1}\, \h\vp (\vx,\eta_{1})\, U_{\eta_{2},\eta_{1}} \quad {\rm and} \quad 
\h\pi (\vx,\eta_{2}) = U_{\eta_{2},\eta_{1}}^{-1}\, \h\pi (\vx,\eta_{1})\, U_{\eta_{2},\eta_{1}}\,\, ? \ee
The counter-intuitive result is that the answer is in the \emph{negative} \cite{cortezetal}. As we will see explicitly in section \ref{s2.3}, unitarity fails even in the simplest, tame example which Parker used in the first investigations of quantum fields in cosmological space-times \cite{parker66}: $a(\eta)= a_{-}$, a constant, in the past until $\eta=\eta_{-}$, then increases monotonically in a smooth fashion till $\eta=\eta_{+}$ and becomes a constant $a(\eta)= a_{+}$ to the future of $\eta_{+}$. This means, the dynamics cannot be transferred to states in the standard fashion; the Schr\"odinger representation does not exist on $\He$. 

Thus, a basic premise of Minkowskian quantum field theories seems to be violated. On the other hand, as we noted above, the covariant approach is in itself complete in the sense that: i) one can describe the dynamics of any observable of direct physical interest; and ii) there is a well-defined, unitary S-matrix in space-times that become asymptotically static in the distant past and future. Our goal is to resolve the apparent tension between the covariant and the canonical descriptions. The analysis will bring out some conceptual subtleties associated with quantum field theory in dynamical space-times that distinguish it from more familiar quantum systems, including quantum fields in Minkowski space-time interacting with time dependent external potentials. Time dependence in space-time geometry has subtle but important conceptual implications that are not shared by field theories on static space-time geometries.

This paper is organized as follows. Because non-unitarity seems counter-intuitive, in section \ref{s2} we will systematically introduce the canonical framework and explain the main result in the context of the simplest time-dependent geometry, that of a FLRW space-time. This discussion serves to make the main issues explicit. In section \ref{s3}, we resolve this tension by proposing a generalization of the notion of unitarity that is appropriate for time dependent geometries. In section \ref{s4} we show that the proposal is compatible, and in fact fits in naturally, with the standard, well-defined S-matrix description in situations in which space-time geometry becomes time independent in the distant past and future. In section \ref{s5} we sketch a generalization of our proposal to arbitrary globally hyperbolic space-times with compact Cauchy surfaces. Section \ref{s6} summarizes the results and discusses conceptual differences between quantum field theory in time dependent space-times and other physical systems.

\section{The canonical framework}
\label{s2}

This section is divided into three parts. In the first we briefly discuss the classical phase space and fix our notation. In the second, we carry out canonical quantization using the algebraic approach where dynamics on the operator algebra $\Acan$ can be specified even before introducing the state space, i.e., without having to decompose the fields into positive and negative frequency parts as in (\ref{fo}). We then introduce the Hilbert space of states and discuss the necessary and sufficient condition for the dynamics defined on $\Acan$ to be unitarily implemented on the Hilbert space. In the third part we show that the condition is not satisfied even in the simplest example mentioned above in section \ref{s1}.

\subsection{The canonical phase space}
\label{s2.1}

Fix a 3-manifold $M$, topologically $\mathbb{T}^{3}$. For the linear  field under consideration, the canonical phase space $\pscan$ consists of pairs $(\vp(\vx), \pi(\vx))\equiv \gamma$ of functions on $M$, equipped with a symplectic structure $\Omega$:
\be \Omega (\g_{1}, \g_{2}) = \sint_{M} \dd^{3}x\, (\vp_{1}\pi_{2} - \vp_{2}\pi_{1})\, .\ee
Using $\Omega$, one can naturally define a set of linear observables $\O_{\g_{o}}$ labeled by points $\g_{o} \equiv (f, g)$ of $\Gamma$ \cite{am2}:
\be \label{classobs}\O_{\g_{o}} (\g) := \Omega(\g_{o}, \g) = \sint_{M} \dd^{3}x\, (f\pi - \vp g)\, .\ee
The vector space they span, together with the one-dimensional space of constant observables on $\pscan$, is closed under Poisson brackets:
\be \{ \O_{\g_{1}},\, \O_{\g_{2}}\} = \Omega (\g_{1},\, \g_{2})\, . \ee
These $\O_{\g_{o}}$ can be regarded as \emph{elementary classical observables} because the full algebra of classical observables is generated by taking linear combinations of their products.  The introduction of elementary observables $\O_{\g}$ is unnecessary in the classical theory but paves the way to quantization from the algebraic perspective, sketched in section \ref{s2.2}.

Let us now consider dynamics. Note that $M$ is an abstract 3-dimensional manifold, rather than a sub-manifold of the 4-manifold $\mathbb{M}$. However, in FLRW backgrounds, given any  value of  the conformal time $\eta_{o}$, there is natural embedding of $M$ into $\mathbb{M}$. This allows us to set an isomorphism $I_{\eta_{o}}$ between the space $\pscov$ consisting of solutions  $\phi(\vx,\eta)$ to the Klein Gordon equation on $\mathbb{M}$ and the canonical phase space $\pscan$:
\ba I_{\eta_{o}}\,\phi(\vx,\eta)  &=& (\vp(\vx), \pi(\vx))\, \in \pscan, \qquad {\rm where} \nonumber\\
\vp(\vx) = \phi(\vx, \eta_{o}), &\,\,& \pi(\vx)
= a^{2}(\eta_{o})\, \big(\partial\phi/\partial\eta\big)\, (\vx,\eta_{o}). \ea
This isomorphism enables us to introduce the natural evolution map  $E_{\eta_{2},\eta_{1}}:\, \pscan \to \pscan$ as $E_{\eta_{2},\eta_{1}} =I_{\eta_2}I^{-1}_{\eta_1}$, such that $\g_{2}:=\, E_{\eta_{2},\eta_{1}}\, \g_{1}$ is the evolution of the phase space point $\g_{1}$ at time $\eta_{1}$, to time $\eta_{2}$. This map $E_{\eta_{2},\eta_{1}}$ in turn provides the dynamical evolution $\Lambda_{\eta_{2},\eta_{1}}$ of observables on $\pscan$. Since under time evolution we have $\Omega(\g_{o}, \g) \mapsto \Omega(\g_{o}, E_{\eta_{2},\,\eta_{1}}\g) =  \Omega(E^{-1}_{\eta_{2},\eta_{1}}\g_{o},\, \g)$, we have
\be \label{auto}\Lambda_{\eta_{2},\eta_{1}}\, \O_{\g} = \O_{\t\g}\qquad {\rm where} \qquad {\t\g} = E^{-1}_{\eta_{2},\eta_{1}}\, \g\, . \ee
$\Lambda_{\eta_{2},\eta_{1}}$ is an automorphism on the space of linear observables in the sense that it preserves their Poisson brackets. It is labelled by \emph{two independent} parameters $\eta_{2}, \eta_{1}$, rather than just their difference $\eta_{2}$-$\eta_{1}$, because the space-time 4-metric $g_{ab}$ on $\mathbb{M}$ ---and hence the Hamiltonian--- is time dependent. Since the elementary observables $\O_{\g}$ generate the full algebra of classical observables, $\Lambda_{\eta_{2},\eta_{1}}$ extends naturally to the full algebra. In the algebraic approach, one regards the (multiplicative and Poisson) algebra of classical observables as primary. The 2-parameter family of automorphisms $\Lambda_{\eta_{2},\eta_{1}}$ specifies dynamics on this algebra.

Now, in stationary space-times dynamics of states is described by a symplectic flow on the phase space, generated by a Hamiltonian vector field. Any such flow naturally induces  a 1-parameter family of automorphisms on the algebra of observables (i.e., mappings that preserve the associative and Poisson algebra structure).  
The space of \emph{all} automorphisms on the algebra of observables is much larger. The 1-parameter families of  automorphisms which are induced from symplectic flows on states constitute a preferred sub-class and are
called \emph{inner automorphisms}. But, as discussed in the previous paragraph, in FLRW space-times we already have dynamical automorphisms $\Lambda_{\eta_{2},\eta_{1}}$ on the canonical algebra. Therefore we are led to ask: Are these automorphisms $\Lambda_{\eta_{2},\eta_{1}}$ \emph{inner}?  That is, do they arise from a symplectic flow on $\pscan$? The answer is in the \emph{negative}, simply because $\Lambda_{\eta_{2},\eta_{1}}$ is a genuinely 2-parameter family of automorphisms, while symplectic flows carry a single parameter. This is of course the standard phenomenon one encounters in any classical system that has a time dependent Hamiltonian. As is well-known, to realize $\Lambda_{\eta_{1},\eta_{2}}$ as an \emph{inner automorphism} in these systems, we need to extend the state space  $\pscan$ to $\psext$ by augmenting it with time:
\be \psext = \pscan \times \mathbb{R}, \qquad \hbox{\rm so that}\,\,\,\, (\gamma, \eta) \in \psext\, . \ee
Then, on each leaf $\pscan^{\eta_{o}}$ of $\psext$ (with $\eta=\eta_{o}$) we have Hamiltonian vector field $X^{\alpha}_{H} = \Omega^{\alpha\beta}\, \partial_{\beta} H(\eta_{o})$, tangential to that leaf, where, for simplicity, we have used Greek indices to label tangent vectors to $\psext$. Dynamics is represented by the flow of the vector field
\be X^{\alpha}_{\rm Dyn} = X^{\alpha}_{H} + \eta^{\alpha} \ee      
on $\psext$, where $\eta^{\alpha}$ is the vector that points in the `time' direction of $\psext$ with affine parameter $\eta$. The first term, $X^{\alpha}_{H}$, on the right side is tangential to each leaf ---and hence `horizontal'--- while the second, $\eta^{\alpha}$, points in the `vertical' direction. This dynamical flow obviously preserves the symplectic structure (whose indices are horizontal). Because each leaf already carries a label $\eta$, the flow generated by $X^{\alpha}_{\rm Dyn}$ provides a well-defined map from the space of observables at any time $\eta_{1}$ to that at time $\eta_{2}$. By construction, it reproduces the action of the automorphism $\Lambda_{\eta_{2},\eta_{1}}$ for all $\eta_{1}, \eta_{2}$. 

To summarize, to make the dynamical map on the space of observables an \emph{inner} automorphism, we have to extend the phase space from $\pscan$ to $\psext$. We have discussed this well known fact explicitly  because, as we will see, the failure of unitary implementability (\ref{unitary}) of dynamics in quantum theory is completely parallel to the impossibility of making $\Lambda_{\eta_{2},\eta_{1}}$ an \emph{inner} automorphism on the standard phase space $\pscan$. In both cases, the dynamical automorphism can be made \emph{inner} by extending the state space.

\subsection {Canonical quantization} 
\label{s2.2}

In the algebraic approach, one begins with operator valued distributions $\h\varphi(\vx), \, \h\pi(\vx)$ on $M$, satisfying the canonical commutation relations 
\be [\h\phi(\vx), \h\pi(\vec{y})] = i\hbar\, \delta(\vx,\, \vec{y})\, . \ee
To display the canonical algebra $\Acan$ generated by them, it is convenient to promote the elementary classical observables $\O_{\g_{o}}$ of (\ref{classobs}) to quantum operators \cite{am2}:
\be \h\O_{\gamma_{o}} := \Omega((f,g),\, (\h\vp,\, \h\pi)) \equiv 
\sint_{M}\, \dd^{3}x\, \big(f(\vx)\h\pi(\vx)\, -\, g(\vx)\h\vp(\vx)\big)\, . \ee  
The canonical quantum algebra $\Acan$ is the free $\star$-algebra generated by these $\h\O_{\gamma}$, subject to the obvious commutation and $\star$ relations:
\be [\h\O_{\gamma_{1}}, \, \h\O_{\gamma_{2}}] = i \hbar\,\,  \Omega  (\g_{1},\, \g_{2}); \quad {\rm and} \quad 
\h\O_{\gamma}^{\star} = \h\O_{\gamma} \ee
for all $\gamma_{1}, \gamma_{2}$ and $\gamma$ of the phase space $\Gamma_{\rm Can}$. The 2-parameter family of dynamical automorphisms $\Lambda_{\eta_{2},\eta_{1}}$ of the classical theory naturally lifts to an automorphisms $\h\Lambda_{\eta_{2},\eta_{1}}$ on $\Acan$:
\be \h\Lambda_{\eta_{2},\eta_{1}} \h\O_{\gamma} := \h\O_{\t\gamma} \qquad {\rm where }\quad \t\gamma = {E^{-1}_{\eta_{2},\eta_{1}}}\, \gamma,\ee
for all $\gamma \in \pscan$. It specifies dynamics on the canonical algebra $\Acan$. Since it is induced directly by the evolution map $E_{\eta_{2},\eta_{1}}$ on the classical phase space $\pscan$, in contrast to the covariant treatment summarized in section \ref{s1}, this evolution does not make any reference to `positive and negative frequency' decomposition, or the Hilbert space of states.

New input ---analogous to the choice of the basis $e_{k}(\eta)$ to decompose fields into `positive and negative frequency parts' in (\ref{fo})--- is required only in the next step, i.e., in the construction of a concrete representation of $\Acan$ by operators on a Hilbert space of states. This step is carried out by first introducing a suitable complex structure $J$ ---i.e., a real linear map satisfying $J^{2}= -I$ on $\pscan$--- which is compatible with the symplectic structure $\Omega$ in the sense that $\Omega(J\g_{1},\, J\g_{2}):= \Omega(\g_{1}, \g_{2})$ and $\Omega(\g, J\g) > 0$ for all non-zero $\g$. Then,  
\be \langle \gamma_{1}, \, \gamma_{2}\rangle \, = \, \f{1}{2\hbar}\, \big(\Omega(\gamma_{1}, J\gamma_{2})\, + i \Omega(\gamma_{1}, \gamma_{2})\big)\ee
is a Hermitian inner product on the complex vector space $(\pscan, J)$ \cite{am1,am2}. (Thus, $(\pscan, \Omega, J)$ is a K\"ahler space.) Then, the positive and negative frequency parts $\gamma^{\pm}$ of $\gamma$ are given by $\gamma^{\pm} = (1/2)(\gamma \mp i J\gamma)$. Thus, while $\g = (\vp,\pi)$ represents a pair of real fields, each of $\g^{\pm}$ represents a pair of complex fields with the property $J\gamma^{\pm} = \pm i\, \g^{\pm}$: on the `positive frequency part' $\g^{+}$, $J$ has the same action as multiplication by $i$ while on the `negative frequency part' $\g^{-}$ it acts as multiplication by $-i$.  
The one-particle Hilbert space $h_J$ is then constructed as the Cauchy completion of $(\pscan, J, \langle \cdot, \cdot \rangle)$, and the Hilbert space $\H_{J}$ is  the symmetric Fock space constructed from $h_J$. One can now introduce creation and annihilation operators on $\H_{J}$ in a standard manner. By their definition, they satisfy the commutation relations
\be [\h{A} (\g_{1}), \h{A}^{\dag}(\g_{2})]\, = \, \langle \g_{1},\, \g_{2}\rangle\, , \ee
and, are complex-linear/anti-linear in their dependence on $\g$:
\be \label{linearity} A^{\dag}(J\g) = i A^{\dag}(\g) \quad {\rm and} \quad A(J\g) = -i A(\g)\, . \ee 
Finally, the abstractly defined field operators $\O_{\gamma}$ are represented in terms of these explicitly defined creation and annihilation operators on $\H_{J}$. The representation map $R_{J}$ is given by
\be \label{rep} R_{J}(\h{\O}_{\g}) = \hbar \big(\h{A}({\g}) + 
\h{A}^{\dag}(\g) \big)\, . \ee
Eqs.\ (\ref{linearity}) and (\ref{rep}) imply 
\be \label{aadag} 
\h{A}(\g) := \f{1}{2\hbar}\,\, R_{J}(\h{\O}_{\g} + i \h{\O}_{J\g}), \qquad{\rm and} \qquad  
\h{A}^{\dag}(\g) := \f{1}{2\hbar}\,\, R_{J}(\h{\O}_{\g} - i \h{\O}_{J\g})\, . \ee
With this kinematic setup at hand, can now discuss dynamics. So far, it is encoded in a 2-parameter family $\h\Lambda_{\eta_{2},\eta_{1}}$ of automorphisms on the abstract algebra $\Acan$. Now that we have a representation $\R_{J}$ of $\Acan$, given any unitary map $U$ on $\H_{J}$, we acquire a distinguished class of automorphisms $\h\Lambda$ on $\Acan$ that satisfy
\be \R_{J}\big(\h\Lambda (\h\O_{\g})\big) = U^{-1}\, \R_{J}(\h{\O}_{\g})\, U 	\, .\ee
 These are called \emph{inner automorphisms}.%
\footnote{Thus, in both classical and quantum theories, \emph{inner} automorphisms on the observable algebra are induced by the structure-preserving isomorphisms on the state space.}
Therefore, we can now ask if the dynamical   automorphisms $\Lambda_{\eta_{2},\eta_{1}}$ is inner. Is so, dynamics would be unitarily implementable on the Hilbert space $\HJ$. This is the case if and only if $\HJ$ admits a 2-parameter family of unitary maps $U_{\eta_{2},\eta_{1}}$ such that
\be \label{unitary1} \langle \chi|\R_{J}\big(\h\Lambda_{\eta_{2},\eta_{1}}\, (\h\O_{\gamma})\big)|\Psi\rangle  = \langle U_{\eta_{2},\eta_{1}}\, \chi|\,  \R_{J}(\h\O_{\gamma})| \, U_{\eta_{2},\eta_{1}} \Psi\rangle, \ee
for all states $\chi, \, \Psi$ in $\H_{J}$ and $\gamma\in \pscan$, or, 
\be \label{unitary2} \R_{J}\big(\h\Lambda_{\eta_{2},\eta_{1}}\, (\h\O_{\gamma})\big)  = U^{-1}_{\eta_{2},\eta_{1}}\, \R_{J}(\h\O_{\gamma}) \, U_{\eta_{2},\eta_{1}}\, . \ee
The operators $U_{\eta_{2},\eta_{1}}$ then provide the evolution of states in the Schr\"odinger picture.

Now, given any complex structure $J$ compatible with the symplectic structure $\Omega$, the evolution map $E_{\eta_{2},\eta_{1}}$ induces a 2-parameter family of complex structures $J_{\eta_{2},\eta_{1}}$ on the phase space $\pscan$ 
\be J_{\eta_{2},\eta_{1}} = E_{\eta_{2},\eta_{1}}\, J\, E_{\eta_{2},\eta_{1}}^{-1}\, ,\ee
each of which is also compatible with $\Omega$. It is well-known (see, e.g. \cite{shale,am2}) that the dynamical automorphism $\h\Lambda_{\eta_{2},\eta_{1}}$ is unitarily implementable if and only if, for all $\eta_{1}, \eta_{2}$,\,the operator $J-J_{\eta_{2},\eta_{1}}$ is Hilbert-Schmidt on the 1-particle Hilbert space $h_J$ 
\be {\rm Tr}_{h_J}\, (J-J_{\eta_{2},\eta_{1}})^{2}  < \infty\, .  \ee
This condition can be restated in a more familiar language as follows. 
Dynamics is unitarily implementable \emph{if and only if the expectation value of the number operator $\h{N}_{\eta_{2},\eta_{1}}$, defined by $J_{\eta_{2},\eta_{1}}$, in the vacuum state $|0\rangle_{J}$ in $\H_{J}$ is finite} for all $\eta_{1},\eta_{2}$. This is the necessary and sufficient condition for the operators $U_{\eta_{2},\eta_{1}}$ in (\ref{unitary2}) to exist.

The counter-intuitive fact is that this condition does not hold for the FLRW space-times under consideration with a generic scale factor $a(\eta)$, irrespective of the initial choice of the complex structure $J$  \cite{cortezetal}. As we will see explicitly in section \ref{s2.3}, this is so even in Parker's tame example mentioned in section \ref{s1}. This means, the dynamics cannot be transferred to states; Schr\"odinger representation does not exist \emph{in the standard sense}.

\subsection{Failure of unitarity: A simple example}
\label{s2.3}

Consider, then, a spatially flat FLRW space-time in which the scale factor $a(\eta)$  is constant $a(\eta)= a_{-}$ in the past until $\eta=\eta_{-}$, then varies smoothly till  $\eta=\eta_{+}$ and becomes a constant $a(\eta)= a_{+}$ to the future of $\eta_{+}$. For our purposes, it suffices to restrict oneself to evolutions from a time $\eta_{1}$ to a time $\eta_{2}$, with $\eta_1<\eta_-$ to $\eta_2>\eta_+$. We will show that, in the generic case, i.e., when  $a_{-} \not= a_{+}$, the dynamical automorphism $\hat{\Lambda}_{\eta_{2}, \eta_{1}}$ introduced in section \ref{s2.2} cannot be unitarily implemented in any Fock representation. The obstruction comes directly from the fact that the underlying space-time geometry is now \emph{dynamical} making $a_{-}$ generically different from $a_{+}$.

Note that the flat space-time regions in the past and the future provide us with two natural quantum representations, the {\it in} and the {\it out}, selected by the Poincar\'e symmetries in the two regions.  We will first prove the result using the {\it in} representation, then extend it to all representations.\\

The setting is provided by the canonical framework of section \ref{s2.2}. Let $|0\rangle _{\rm in}$ be the {\it in} vacuum defined by $J_{\rm in}$, and $\h{N}_{\eta_{2},\eta_{1}}$ the number operator in the representation defined by $J_{\eta_{2},\eta_{1}}=  E_{\eta_{2},\eta_{1}} J_{\rm in} E_{\eta_{2},\eta_{1}}^{-1}$. Given any `positive frequency' basis $\{\gamma^{\rm in}_{\vec{k}}(\vec{x})\}$ in the 1-particle Hilbert space $h_{J_{\rm in}}$  defined by the complex structure $J_{\rm in}$, the set $\{E_{\eta_{2},\eta_{1}} \gamma^{\rm in}_{\vec{k}}(\vec{x})\}$ is a complete basis of the `positive frequency' subspace of $h_{J_{\eta_2,\eta_1}}$. Completeness of the {\it in} basis ensures that the two sets of vectors are related by a Bogoliubov transformation 
\be \label{bog}   E_{\eta_{2},\eta_{1}} \gamma^{\rm in}_{\vec{k}}(\vec{x})= \alpha_{\vec{k}} \, \gamma^{\rm in}_{\vec{k}} ({\vec x})+\beta_{\vec{k}} \, \gamma^{{\rm in}\, \star}_{\vec{k}} ({\vec x})\,,\ee 
for some coefficients $\alpha_{\vec{k}}$ and $\beta_{\vec{k}}$ which we want to determine. The normalization of the basis vectors implies $|\alpha_{\vec{k}} |^2-|\beta_{\vec{k}}|^2=1$. The criterion for the existence of unitary evolution operators $U_{\eta_1\eta_2}$ was given at the end of section \ref{s2.2}. For the {\it in} representation, it translates to

\be \label{cond} _{\rm in}\langle 0|\h{N}_{\eta_{2},\eta_{1}}|0\rangle _{\rm in}=\sum_{\vec{k}} |\beta_{\vec{k}}|^2 <\infty \, , \ee
Different  choices of basis in  $h_{J_{\rm in}}$ will change the coefficients $\beta_{\vec{k}}$ at most by a phase factor. Therefore, the summability of $|\beta_{\vec{k}}|^2$ appearing in condition  (\ref{cond}) does not depend on the specific choice.  
 
To check whether condition  (\ref{cond}) is satisfied, it is helpful to also consider the positive frequency basis  $\gamma^{\rm out}_{\vec{k}}$ defined by the complex structure $J_{\rm out}$. We will take advantage of the following two results concerning the relation between $\gamma^{\rm out}_{\vec{k}}$ and $\gamma^{\rm in}_{\vec{k}}$: 

{\bf Lemma 1:} {\it The complex Bogoliubov coefficients $\sigma^{(1)}_{\vec{k}}$ and  $\sigma^{(2)}_{\vec{k}}$, defined via 
\be \label{sigmas} \gamma^{\rm out}_{\vec{k}} ({\vec x}) = \sigma^{(1)}_{\vec{k}} \, \gamma^{\rm in}_{\vec{k}} ({\vec x}) + \sigma^{(2)}_{\vec{k}}\, \gamma^{{\rm in}\, \star}_{\vec{k}} ({\vec x})\, , \ee
 have the following large-momentum asymptotic behavior 
\be \label{assympt-sigmas} \sigma^{(1)}_{\vec{k}}=\frac{1}{2}\left[\frac{a_-^2+a_+^2}{a_+\, a_-}\right]+\, \mathcal{O}\left(\frac{m^2}{k^{2}}\right)\, ; \hspace{0.3cm} \sigma^{(2)}_{\vec{k}}=\frac{1}{2}\left[\frac{a_-^2-a_+^2}{a_+\, a_-}\right]+\, \mathcal{O}\left(\frac{m^2}{k^{2}}\right)\, ,\hspace{0.3cm}  {\rm when}\, \,  k=|\vec{k}|\to \infty \ee
}

{\it Proof:}
Since $J_{\rm in}$ and $J_{\rm out}$ are the standard  flat space complex structure in the past and future, respectively, natural bases of `positive frequency' modes are given by
\be  \gamma_{\vec{k}}^{\rm in}(\vec{x})=(e^{\rm in}_{\vec{k}}\, e^{i\vec{k}\cdot\vec{x}},\,\, f^{\rm in}_{\vec{k}}\, e^{i\vec{k}\cdot\vec{x}}) \, , \hspace{1cm}  \gamma_{\vec{k}}^{\rm out}(\vec{x})=(e^{\rm out}_{\vec{k}}\, e^{i\vec{k}\cdot\vec{x}},\,\,f^{\rm out}_{\vec{k}}\, e^{i\vec{k}\cdot\vec{x}})\, , \ee
where
\be \nonumber e^{\rm in}_{\vec{k}}=\frac{1}{a_- \sqrt{2 w_{\rm in}}}\, ; \hspace{0.3cm} f^{\rm in}_{\vec{k}}=a_- \left(\frac{-i w_{\rm in}}{ \sqrt{2 w_{\rm in}}}\right) \,; \hspace{0.3cm}  e^{\rm out}_{\vec{k}}=\frac{1}{a_+ \sqrt{2 w_{\rm out}}}\, ;\hspace{0.3cm} f^{\rm out}_{\vec{k}}=a_+ \left(\frac{-i w_{\rm out}}{ \sqrt{2 w_{\rm out}}} \right)\, \ee
with $w_{\rm in}=\sqrt{k^2+m^2\,  a_-^2}$, $w_{\rm out}=\sqrt{k^2+m^2\, a_+^2}$. Note that these modes provide Cauchy data for standard positive frequency plane-waves in flat space.  Equations (\ref{sigmas}) provide two algebraic relations  for $\sigma^{(1)}_{\vec{k}}$ and $\sigma^{(2)}_{\vec{k}}$, for which the solutions are 

\be \sigma^{(1)}_{\vec{k}}=\frac{1}{2}\frac{a_-^2\, w_{\rm in}+a_+^2\, w_{\rm out}}{a_+ a_- \sqrt{w_{\rm out}w_{\rm in}}}\, ; \hspace{0.5cm} \sigma^{(2)}_{\vec{k}}=\frac{1}{2}\frac{a_-^2\, w_{\rm in}-a_+^2\, w_{\rm out}}{a_+ a_- \sqrt{w_{\rm out}w_{\rm in}}} \, .\ee
Expanding for large $k$ one obtains the asymptotic expressions (\ref{assympt-sigmas}). \,\, $\Box$\\

{\bf Lemma 2:} {\it The complex functions $\lambda^{(1)}_{\vec{k}}$ and  $\lambda^{(2)}_{\vec{k}}$, defined via 

\be \label{lambdas}  E_{\eta_2,\eta_1} \gamma^{\rm in}_{\vec{k}} = \lambda^{(1)}_{\vec{k}} \, \gamma^{\rm out}_{\vec{k}}+\lambda^{(2)}_{\vec{k}}\, \gamma^{{\rm out}\, \star}_{\vec{k}}\,\ee
have the following asymptotic  behavior 

\be \label{assympt-lambdas}  \lambda^{(1)}_{\vec{k}}\to 1\, , \hspace {0.5cm} {\rm and} \quad \lambda^{(2)}_{\vec{k}}\to 0 \, \hspace{0.5cm}
{\rm as}\quad  k\to \infty\, \ee
faster than any power of $k^{-1}$. }

 This is a special case of a theorem due to Kulsrud \cite{Kulsrud}. (The proof relies on the use of adiabatic invariants for the time dependent harmonic oscillator, (see also \cite{parker-book}).)\,\, $\Box$\\
 
With these two results in hand, we are ready to proof the following result.
 
\begin{theorem} The dynamical automorphism $\hat{\Lambda}_{\eta_{2}, \eta_{1}}$ can be unitarily implemented  in the  {\it in}-Fock representation if and only if $a_- = a_+$.
\end{theorem}
 
{\it Proof:} 
We only need to compute the asymptotic behavior of the coefficients $\beta_{\vec{k}}$ defined in equation (\ref{bog}).  We have:
\ba    E_{\eta_2,\eta_1}\, \gamma^{\rm in}_{\vec{k}}&=& ( \lambda^{(1)}_{\vec{k}} \, \gamma^{\rm out}_{\vec{k}}+\lambda^{(2)}_{\vec{k}}\, \gamma^{{\rm out}\, \star}_{\vec{k}})\nonumber \\& =&  \lambda^{(1)}_{\vec{k}} \, (\sigma^{(1)}_{\vec{k}} \, \gamma^{\rm in}_{\vec{k}}+\sigma^{(2)}_{\vec{k}}\, \gamma^{{\rm in}\, \star}_{\vec{k}})+\lambda^{(2)}_{\vec{k}}\, (\sigma^{(1)\, \star}_{\vec{k}} \, \gamma^{{\rm in}\, \star}_{\vec{k}}+\sigma^{(2)\, \star}_{\vec{k}}\, \gamma^{\rm in}_{\vec{k}})\nonumber\\&=&(\lambda^{(1)}_{\vec{k}} \, \sigma^{(1)}_{\vec{k}} +\lambda^{(2)}_{\vec{k}}\, \sigma^{(2)\, \star}_{\vec{k}})\,  \gamma^{{\rm in}}_{\vec{k}} +(\lambda^{(1)}_{\vec{k}} \, \sigma^{(2)}_{\vec{k}} +\lambda^{(2)}_{\vec{k}}\, \sigma^{(1)\, \star}_{\vec{k}})\,  \gamma^{{\rm in\, \star}}_{\vec{k}} 
\, \ea
where  equation (\ref{lambdas}) has been used in the first equality, and equation (\ref{sigmas}) in the second one. Comparing this equation with (\ref{bog}), we obtain

\be \alpha_{\vec{k}}= \lambda^{(1)}_{\vec{k}} \, \sigma^{(1)}_{\vec{k}} +\lambda^{(2)}_{\vec{k}}\, \sigma^{(2)\, \star}_{\vec{k}} \, ; \hspace{1cm} \beta_{\vec{k}}= \lambda^{(1)}_{\vec{k}}\, \sigma^{(2)}_{\vec{k}}+ \lambda^{(2)}_{\vec{k}}\sigma^{(1)\, \star}_{\vec{k}} \, .\ee
The asymptotic expressions (\ref{assympt-sigmas}) and (\ref{assympt-lambdas}) then imply 
\be \label{assymp} \alpha_{\vec{k}}=\frac{1}{2}\left[\frac{a_-^2+a_+^2}{a_+\, a_-}\right]+\,  \mathcal{O}\left(\frac{m^2}{k^2}\right)\, ;\hspace{0.5cm} \beta_{\vec{k}}=\frac{1}{2}\left[\frac{a_-^2-a_+^2}{a_+\, a_-}\right]+\,  \mathcal{O}\left(\frac{m^2}{k^2}\right)\, ,\hspace{0.3cm}  {\rm when}\, \,  k\to \infty \, . \ee
Clearly, the $\beta_{\vec{k}}$ coefficients are square-summable, i.e.\  $\sum_{\vec{k}} |\beta_{\vec{k}}|^2<\infty$, if and only if $a_-=a_+$.
\footnote{We assume that $a(\eta)$ is a smooth function. In the case $a_-=a_+$, for $\beta_{\vec{k}}$ to be square-summable one only needs $a(\eta)$ to be $C^{2}$.}  
We conclude that the condition  (\ref{cond}) is  satisfied, and therefore the dynamical automorphism can be unitarily implemented  in the  {\it in}-Fock representation, only in the  special situation $a_-=a_+$. \,\,$\Box$

To summarize, on the canonical phase space $\pscan$, one can introduce three complex structures, $J_{\rm in}$ and $J_{\rm out}$ selected by the past and future Poincar\'e symmetries and $J_{\eta_{2},\eta_{1}} = E_{\eta_{2},\eta_{1}}\,J_{\rm in}\,E^{-1}_{\eta_{2},\eta_{1}}$ obtained by dynamically evolving $J_{\rm in}$ to the future. Lemma 1 says that, if $a_{-} \not= a_{+}$, then $(J_{\rm in} - J_{\rm out})$ fails to be Hilbert-Schmidt. Lemma 2 says that $(J_{\rm out} - J_{\eta_{2},\eta_{1}})$ \emph{is} Hilbert Schmidt. Therefore, $(J_{\rm in} - J_{\eta_{2},\eta_{1}})$ fails to be Hilbert-Schmidt, whence $\h\Lambda_{\eta_{2},\eta_{1}}$ can not be unitarily implemented on $\H_{J_{\rm in}}$. Thus, of the three representation of the \emph{canonical} algebra $\Acan$, only the ones determined by $J_{\eta_{2},\eta_{1}}$ and $J_{\rm out}$ are unitarily equivalent. We will return to this interplay in section \ref{s4}. \\   

A natural question now arises: Is there any other representation, or equivalently any other complex structure $\tilde J$, in which dynamics is unitary for generic $a(\eta)$ with our asymptotic behavior? The answer is also in the negative, as we now show. Let $\tilde \gamma_{\vec{k}}(\vec{x})$ be the basis of the `positive frequency' subspace $h_{\tilde J}$ associated to $\tilde J$. Using the same argument as above, dynamics will be unitary if  $\sum_{\vec{k}}|\tilde \beta _{\vec{k}}|<\infty$, where the Bogoliubov coefficients $\tilde \beta_{\vec{k}}$ are defined by
\be \label{tildeab} E_{\eta_2,\eta_1}  \tilde \gamma_{\vec{k}}= \tilde \alpha_{\vec{k}} \,  \tilde \gamma_{\vec{k}}+\tilde \beta_{\vec{k}}  \, \tilde \gamma^{ \star}_{\vec{k}} \, . \ee
It is useful to characterize the freedom in the choice of  $\tilde J$ through the two real functions $r_{\vec{k}}$ and $\theta_{\vec{k}}$ defined by
\be \label{tildein} \tilde \gamma_{\vec{k}}(\vec{x}) = (r_{\vec{k}}^2+1)^{1/2}e^{i \theta_{\vec{k}}} \, \gamma^{\rm in}_{\vec{k}} (\vec{x})+r_{\vec{k}}\, \gamma^{\rm in\, \star}_{\vec{k}} (\vec{x})\, , \ee
for al $\vec{k}$. There is a one-to-one relation between families of pairs $\{r_{\vec{k}},\theta_{\vec{k}}\}$ for every $\vec{k}$ and complex structures $\tilde J$. Therefore, the question we want to answer is if there exist at least one family $\{r_{\vec{k}},\theta_{\vec{k}}\}$ for which the coefficients $\tilde \beta_{\vec{k}}$  are square summable. We will proceed by writing $\tilde \beta_{\vec{k}}$  in terms of $r_{\vec{k}}$,  $\theta_{\vec{k}}$,  $\alpha_{\vec{k}}$ and $\beta_{\vec{k}}$, where $\alpha_{\vec{k}}$ and $\beta_{\vec{k}}$ were defined in (\ref{bog}). This can be done as follows 
\ba E_{\eta_2,\eta_1}\, \tilde \gamma_{\vec{k}}&=& (r_{\vec{k}}^2+1)^{1/2}e^{i \theta_{\vec{k}}} \, (E_{\eta_2,\eta_1} \gamma^{\rm in}_{\vec{k}})+r_{\vec{k}}\, (E_{\eta_2,\eta_1}\gamma^{\rm in\, \star}_{\vec{k}}) \\& =&  (r_{\vec{k}}^2+1)^{1/2}e^{i \theta_{\vec{k}}} \, ( \alpha_{\vec{k}} \, \gamma^{\rm in}_{\vec{k}}+\beta_{\vec{k}} \, \gamma^{{\rm in}\, \star}_{\vec{k}})+r_{\vec{k}}\, ( \alpha^{\star}_{\vec{k}} \, \gamma^{\rm in\, \star }_{\vec{k}}+\beta^{\star}_{\vec{k}} \, \gamma^{{\rm in}}_{\vec{k}})\nonumber \\& =& \left[(r_{\vec{k}}^2+1)^{1/2}e^{i \theta_{\vec{k}}}\, \alpha_{\vec{k}}+r_{\vec{k}} \, \beta^{\star}_{\vec{k}}\right]\, \gamma^{{\rm in}}_{\vec{k}}+ \left[(r_{\vec{k}}^2+1)^{1/2}e^{i \theta_{\vec{k}}}\, \beta_{\vec{k}}+r_{\vec{k}} \, \alpha^{\star}_{\vec{k}}\right]\, \gamma^{{\rm in\, \star}}_{\vec{k}}\, .  \nonumber \ea
We have used (\ref{tildein}) and the linearity of $E_{\eta_1,\eta_2}$ in the first equality, and (\ref{bog}) in the second. Comparing with equations (\ref{tildeab}) and (\ref{tildein})  we find

\be \tilde \beta_{\vec{k}}=-r_{\vec{k}} \, \left[(r_{\vec{k}}^2+1)^{1/2}e^{i \theta_{\vec{k}}}\, \alpha_{\vec{k}}+r_{\vec{k}} \, \beta^{\star}_{\vec{k}}\right]+\,(r_{\vec{k}}^2+1)^{1/2}e^{i \theta_{\vec{k}}}\,  [(r_{\vec{k}}^2+1)^{1/2}e^{i \theta_{\vec{k}}}\, \beta_{\vec{k}}+r_{\vec{k}} \, \alpha^{\star}_{\vec{k}}] \, .\ee
Now, the  behavior of $\beta_{\vec{k}}$ for large $k$ follows from the asymptotic behavior of $\alpha_{\vec{k}}$ and $\beta_{\vec{k}}$ given in equation (\ref{assymp})

\be \tilde \beta_{\vec{k}}= \left[r_{\vec{k}}^2(-1+e^{-i 2\theta_{\vec{k}}})+e^{-i 2\theta_{\vec{k}}}\right]\, \left(\frac{1}{2}\frac{a_-^2-a_+^2}{a_+a_-}\right)+\mathcal{O}\left(\frac{m^2}{k^2}\right) \, .\ee
 There exist no positive real function $r_{\vec{k}}$ and angle  $\theta_{\vec{k}}$ such that  $\sum_{\vec{k}} |\tilde \beta_{\vec{k}}|^2<\infty$, unless $a_-=a_+$ (in which case $\tilde \beta_{\vec{k}}=0$ for all $\vec{k}$). Thus, we have arrived at our final result:
\begin{theorem} The time evolution from $\eta_1<\eta_-$ to $\eta_2>\eta_+$ cannot be implemented by a unitary operator in any Fock space $\mathcal{H}_J$, irrespective of the choice of the complex structure $J$ on the canonical phase space $\pscan$.
\end{theorem}

\section{Resolution of the tension}
\label{s3}

We now wish to resolve the tension between the self-contained and complete description in the Heisenberg picture used in the covariant formulation (summarized in section \ref{s1}), and the failure of unitarity in the canonical picture (discussed in section \ref{s2}). Therefore, we begin in section \ref{s3.1} with a statement of the relation between the structures used in the two frameworks. In section \ref{s3.2} we resolve the apparent tension by generalizing the standard formulation of unitary dynamics in the canonical formulation  and discuss several conceptual issues related to this resolution.  In section \ref{s3.3} we make this resolution explicit through an example from the cosmological perturbation theory.

\subsection{Relation between the covariant and canonical theories}
\label{s3.1}

Let us begin by reconstructing the covariant theory using an algebraic approach similar to the one we used in the canonical theory in section \ref{s2}. Consider an abstractly defined operator valued distribution $\h\phi(\vx,\eta)$ on $\mathbb{M}$, satisfying the field equation, $\star$-relations and the commutation relations
\ba (\Box - m^2)\, \h\phi (\vx,\eta) = 0, &\quad& \h\phi^{\star}(\vx, \eta) = \h\phi (\vx,\eta),\qquad {\rm and} \nonumber\\ 
\big[\h{\phi}(\vx_{1}, \eta_{1}),\,\, \h{\phi}(\vx_{2}, \eta_{2}) \big]  &=& i\hbar\,   \big (G_{\rm ret} (\vx_{1}, \eta_{1};\, \vx_{2}, \eta_{2}) - G_{\rm adv} (\vx_{1}, \eta_{1};\, \vx_{2}, \eta_{2})\big)\, ,  \ea
where $G_{\rm ret}$ and $G_{\rm adv}$ are the retarded and advanced Green functions. The covariant $\star$-algebra $\Acov$ is generated by operators $\h\phi(f) := \int_{\mathbb{M}} {\rm d}^{4}V\, \h\phi(\vx, \eta) f(\vx,\eta)$ obtained by smearing $\hat\phi(\vx, \eta)$ with test fields $f(\vx, \eta)$ on $\mathbb{M}$. To find a representation of this algebra, one normally introduces a  `positive frequency' basis $e_{k}(\eta)$ satisfying (\ref{e1}) and (\ref{e2}). This basis defines a complex structure $\J$ on the covariant phase space $\pscov$ consisting of real classical solutions $\phi(\vx,\eta)$ to the field equation on $\mathbb{M}$ as follows. Given the expansion 
\be \phi(\vx,\eta) = \frac{1}{V_{0}}\, \sum_{\vk}\, [e_{k}(\eta) {A}_{\vk}\, + \, e^{\star}_{k}(\eta) A^{\star}_{-\vk} ]\,\, e^{i\vk\cdot\vx}\, , 
\ee
of $\phi(\vx,\eta)$ in terms of these basis vectors (with complex coefficients $A_{k}$), the real solution $(\J\, \phi)(\vx,\eta)$ is given by%
\footnote{Properties of basis functions ensure that the complex structure is automatically compatible with the natural symplectic structure $\Omega_{\rm cov}$ on $\pscov$: $\Omega_{\rm cov}(\phi_{1}, \phi_{2}) = \int (\phi_{1}\nabla_{a}\phi_{2} -  \phi_{2}\nabla_{a}\phi_{1})\, \dd S^{a}$, where the integral is performed on any Cauchy slice of $\mathbb{M}$.}
\be \J\,\phi(\vx,\eta) = \frac{1}{V_{0}}\, \sum_{\vk}\, [i\,e_{k}(\eta) A_{\vk}\, - i \, e^{\star}_{k}(\eta) A^{\star}_{-\vk} ]\,\, e^{i\vk\cdot\vx}\, . \ee
In fact, the invariant content in the choice of the basis $e_{k}(\eta)$ is captured precisely in the complex structure $\J$ it determines: two choices $e_{k}(\eta)$ and $\t{e}_{k}(\eta)$ lead to the same vacuum $|0\rangle$ (and hence the same Fock representation of $\Acov$) if and only if they define the same $\J$. Therefore, it is more appropriate to denote the representation space by $\H_{\J}$ (rather than $\H_{\{e_{k}\}}$). Thus, from the algebraic viewpoint, the new input needed in the passage from the classical phase space $\pscov$ to the quantum theory $(\Acov, \H_{\J})$ is, again, an appropriate complex structure $\J$.\\

Let us now relate the structures of the covariant and canonical formulations. Fix a $\J$ on $\pscov$ and the corresponding representation of $\Acov$ on the Fock space $H_{\J}$ it determines (through (\ref{fo})). Recall from section \ref{s2.1} that for any given $\eta_{o}$, there is a natural isomorphism $I_{\eta_{o}}$ from $\pscov$ to $\pscan$. Therefore, the complex structure $\J$ on $\pscov$ induces a 1-parameter family of complex structures $J_{\eta}$ on $\pscan$ via
\be \label{Jeta} J_{\eta} := I_{\eta}\, \J\, I_{\eta}^{-1}\, .     \ee
By their definition, these complex structures on $\pscan$ are related to one another by
\be \label{relation}
 J_{\eta_{1}}  = E^{-1}_{\eta_{2},\eta_{1}}\, J_{\eta_{2}}\, E_{\eta_{2},\eta_{1}}  \ee
where, as before, $E_{\eta_{2},\eta_{1}}=I_{\eta_2}I^{-1}_{\eta_1}$ is map on $\pscan$ that evolves states at time $\eta_{1}$ to time $\eta_{2}$. Each $J_{\eta}$ is compatible with the symplectic structure $\Omega$ on $\pscan$. Therefore, the construction discussed in section \ref{s2.2} now provides us with a \emph{one parameter family} of representations $R_{J_{\eta}}$ on Hilbert spaces $\H_{J_{\eta}}$ of the canonical $\star$-algebra $\Acan$. 

To summarize, each choice of an adiabatic basis $e_{k}(\eta)$ satisfying (\ref{e1}) and (\ref{e2}) provides us a representation of the abstract operator algebra $\Acov$ of the covariant theory. On the canonical phase space $\pscan$, this basis induces a 1-parameter family of (`adiabatically regular') complex structures $J_{\eta}$, thereby providing us with a 1-parameter family $R_{J_{\eta}}$ of representations of the canonical algebra $\Acan$ by operators on Fock spaces $\H_{J_{\eta}}$. Put differently, there is no 1-1 relation between the textbook covariant quantum theory summarized in section \ref{s1} and the canonical quantum theory discussed in section \ref{s2} because the quantum theory based on the covariant phase space $\pscov$ does \emph{not} provide us with a single Hilbert space $\H$ of states and a representation $R$ of the canonical algebra $\Acan$. There is, however, a clearcut relation if we use the \emph{extended} phase space $\psext$ in place of $\pscan$. For, the covariant theory provides us with a Hilbert space $\H_{J_{\eta}}$ for each horizontal leaf $\psext^{\eta}$ of $\psext$ and a representation $R_{J_{\eta}}$ of the canonical algebra $\Acan$ by operators on $\H_{J_{\eta}}$. In this precise sense, the covariant theory provides us with a specific quantization of the extended phase space $\psext$ rather than the canonical phase space $\pscan$.\\

\emph{Remark:} What would happen if we considered a \emph{static} space-time in place of FLRW? In the classical theory, one would have a time independent Hamiltonian selected by the presence of the isometry and hence dynamics is represented by a Hamiltonian flow (inner automorphism) on the canonical phase space $\pscan$; we do not have to construct $\psext$. In the quantum theory, the presence of the isometry enables us to select a \emph{preferred} complex structure $\J$ on $\pscov$ (assuming the norm of the Killing field remains strictly bounded away from zero) \cite{am1}. Furthermore, the complex structures $J_{t}$ on $\pscan$, associated with the leaves of the static foliation, all agree (under the \emph{kinematical} identification). Therefore, in this case, one has a single representation of the canonical algebra $\Acan$ on a Fock space $\H_{J}$ on which dynamics is unitarily represented. Thus, the situations in the classical and quantum domains are again parallel, but they differ from the corresponding situations in the FLRW space-times where the space-time geometry is dynamical.

\subsection{Dynamics and generalized unitarity}
\label{s3.2}

 This subsection is divided into two parts. In the first, we extend the notion of unitarity and show that the Heisenberg dynamics on $\Acan$ is in fact unitarily implemented in this extended sense. In the second, we  discuss a subtlety that arises because of the time dependence of the underlying geometry.

\subsubsection{Schr\"odinger evolution on the extended state space}
\label{s3.2.1}

Let us now revisit the question of unitary implementation of the 2-parameter family of automorphisms $\hat\Lambda_{\eta_{2},\eta_{1}}$ on $\Acan$. The relation between the covariant and canonical descriptions spelled out in section \ref{s3.1} provides a more general framework to formulate and analyze this issue. Now that we have a  1-parameter family of representations $(R_{J_{\eta}}, \H_{J_{\eta}})$ of $\Acan$, in place of (\ref{unitary1}), we are now led to ask: \emph{Given $\eta_{2}, \eta_{1}$, does there exist a unitary map} 
\be  U_{\eta_{2},\eta_{1}}\,\, :\,\, \H_{J_{1}} \longrightarrow  \H_{J_{2}} \ee   
\emph{such that}
\be \label{implementation1} \langle \chi_{1}|\R_{J_{1}}\big(\h\Lambda_{\eta_{2},\eta_{1}}\, (\h\O_{\gamma})\big)|\Psi_{1}\rangle  = \langle U_{\eta_{2},\eta_{1}}\, \chi_{1}|\,  \R_{J_{2}}(\h\O_{\gamma})| \, U_{\eta_{2},\eta_{1}} \Psi_{1}\rangle \, ?  \ee
Here and in what follows we have set $J_{1} = J_{\eta_{1}}$ and $J_{2} = J_{\eta_{2}}$ for notational simplicity. The left side of (\ref{implementation1}) provides us the evolution of the matrix element in the Heisenberg picture in which the operator has evolved to time $\eta_{2}$ but states are frozen at the initial time $\eta_{1}$, while on the right side the states evolve via unitary maps $U_{\eta_{2},\eta_{1}}$ while the operator is frozen at the initial time $\eta_{1}$. In terms of operators, then, we are led to generalize (\ref{unitary2}) to seek operators $U_{\eta_{2},\eta_{1}}$ satisfying 
\be \label{implementation2}  R_{J_{1}}\,(\hat{\Lambda}_{\eta_{2},\eta_{1}}\,\h{\O_{\gamma}}) \,   \,=\, U^{{-1}}_{\eta_{2},\eta_{1}}\, (R_{J_{2}} (\h{\O_{\gamma}}))\, U_{\eta_{2},\eta_{1}}\,  \ee
for all $\g$. In the extended framework introduced in section \ref{s3.1}, this is the natural formulation of the problem. Indeed, it is completely analogous to the formulation in the classical theory, where we ask if there is a symplectic map from the leaf $\psext^{\eta_{1}}$ of the extended phase space, to its leaf $\psext^{\eta_{2}}$, which implements the dynamical automorphism $\Lambda_{\eta_{2},\eta_{1}}$. That is, in both cases, the question is whether the dynamical automorphism on the observable algebra can be made \emph{inner} via mappings on the extended state space from states at time $\eta_{1}$ to the states at time $\eta_{2}$.

To answer this question, we use two facts. First, the definition (\ref{auto}) of the dynamical automorphisms provides an explicit expression
of  $\h\Lambda_{\eta_{2},\eta_{1}} \h\O_{\gamma}$ that appears on the left side of (\ref{implementation2}):
\be \h\Lambda_{\eta_{2},\eta_{1}} \h\O_{\gamma} := \h\O_{\t\gamma} \qquad {\rm where }\quad \t\gamma =  E^{-1}_{\eta_{2},\eta_{1}}\, \gamma\, .\ee
Second, we use expressions (\ref{aadag}) of the annihilation and creation operators, $\h{A}({\gamma}),\, \h{A}^{\dag}({\gamma})$ 
\be \label{aadag2} 
\h{A}({\g}) := \f{1}{2\hbar}\,\, R_{J}(\h{\O}_{\g} + i \h{\O}_{J\g}), \qquad{\rm and} \qquad  
\h{A}^{\dag}({\g}) := \f{1}{2\hbar}\,\, R_{J}(\h{\O}_{\g} - i \h{\O}_{J\g})\, . \ee
These two inputs enable us to extract from (\ref{implementation2}) the relation between the annihilation operators $\h{A}_{2}$ on $\H_{J_{2}}$ and the creation and annihilation operators $\h{A}_{1}^{\dag}$ and $\h{A}_{1}$ on $\H_{J_{1}}$ that the desired unitary map $U_{\eta_{2},\eta_{1}}$ must yield:
\be \label{equiv1} -2i \h{A}_{1}(E^{-1}_{\eta_{2},\eta_{1}}\g) = U^{-1}_{\eta_{2},\eta_{1}}\,\,\Big(\h{A}_{2}([E_{\eta_{2},\eta_{1}}\,J_{1}\, E^{-1}_{\eta_{2},\eta_{1}}+J_{2}]\g) \, +\, \h{A}_{2}^{\dag} ([E_{\eta_{2},\eta_{1}}J_{1}E^{-1}_{\eta_{2},\eta_{1}} - J_{2}]\g) \Big)\,\,U_{\eta_{2},\eta_{1}}\, ,  \ee
where we have used the consequence,  $\h{A}({J\g}) = -i \h{A}({\g})$ and $\h{A}^{\dag}({J\g}) = i \h{A}^{\dag} ({J\g})$, of (\ref{aadag2}). Applying this equation to the vacuum state $|0_{1}\rangle$ in $\H_{J_{1}}$ and setting $U_{\eta_{2},\eta_{1}} |0_{1}\rangle = |\Psi_{2}\rangle\in 
\H_{J_{2}}$, we find
\be \label{equiv2}0\,=\,\big(\h{A}_{2}([ E_{\eta_{2},\eta_{1}}\,J_{1}\, E^{-1}_{\eta_{2},\eta_{1}} +J_{2}]\g) \, +\, \h{A}_{2}^{\dag} ([E_{\eta_{2},\eta_{1}}J_{1}E_{\eta_{2},\eta_{1}} - J_{2}]\g)\big) |\Psi_{2}\rangle, \quad \forall \gamma\in \pscan\, . \ee 
Now, the argument used in the standard analysis \cite{shale,am2} of the unitary equivalence of the two Fock representations of $\Acan$ implies that the desired unitary maps $U_{\eta_{2},\eta_{1}}$ exist if and only if  there is a normalizable state $|\Psi_{2}\rangle\in \H_{J_{2}}$ satisfying (\ref{equiv2}). This in turn is the case if and only if 
\be \label{equiv} [E_{\eta_{2},\eta_{1}}\, J_{1}\, E^{-1}_{\eta_{2},\eta_{1}} - J_{2}]\quad {\hbox{\rm is Hilbert-Schmidt}} \ee
(on the 1-particle Hilbert space $h_{J_{2}}$). But recall that, since the complex structures $J_{\eta}$ are induced on $\pscan$ by a single (adiabatically regular) complex structure $\J$ on $\pscov$, we have $J_{2} = E_{\eta_{2},\eta_{1}}\, J_{1}\, E^{-1}_{\eta_{2},\eta_{1}}$. Therefore the Hilbert-Schmidt requirement is trivially satisfied and dynamics is unitarily implemented in the precise sense spelled out in  (\ref{implementation2}).\\  

\emph{Remark:} Because the Hilbert-Schmidt requirement is trivially satisfied whenever the family $J_{\eta}$ is induced by a single complex structure $\J$ on $\pscov$, one may think that the resulting dynamics in the generalized Schr\"odinger picture is also trivial in some sense. We want to emphasize that \emph{this is not the case}; in this Schr\"odinger picture, the time evolution of states faithfully mirrors the Heisenberg dynamics of operators as usual. This is simply because, by construction, the unitary operators $U_{\eta,\eta_{0}}$ satisfy (\ref{implementation2}):
\be \langle \chi_{0}|\R_{J_{\eta_{0}}}\Big(\h\Lambda_{\eta,\eta_{0}}\, (\h\O_{\gamma})\big)|\Psi_{0}\rangle  = \langle U_{\eta,\eta_{0}}\, \chi_{0}|\,  \R_{J_{\eta}}(\h\O_{\gamma})| \, U_{\eta,\eta_{0}} \Psi_{0}\rangle . \ee
The left side is manifestly $\eta$-\emph{dependent} and provides the evolution of the matrix elements of the  Heisenberg operators in the `initial' Hilbert space $\H_{J_{\eta_{0}}}$. The right side refers to the Schr\"odinger picture. The equality \emph{guarantees} that the evolution of states, $|\Psi_{0}\rangle  \to U_{\eta,\eta_{0}}|\Psi_{0}\rangle$ and  $|\chi_{0}\rangle  \to U_{\eta,\eta_{0}}|\chi_{0}\rangle$, in this picture is both non-trivial and correct. In the next subsection we will make these considerations explicit using 
2-point functions.

To summarize, if we equip $\pscan$ with a single complex structure $J$ then, as we saw in section \ref{s2.3}, the dynamical automorphisms $\h\Lambda_{\eta_{2},\eta_{1}}$ on $\Acan$ cannot be made `inner' ---i.e., unitarily implemented--- on the Fock space $\H_{J}$ for a generic FLRW space-time. But in this case, \emph{already in the classical theory}, the dynamical automorphisms $\Lambda_{\eta_{2},\eta_{1}}$ on the classical observable algebra cannot be made \emph{inner} on the phase space $\pscan$. To make them \emph{inner}, one has to extend $\pscan$ to $\psext$. Therefore, in the quantum theory, it is natural to first assign a complex structure $J_{\eta}$ to each leaf $\psext^{\eta}$ of $\psext$ and construct the corresponding representations $(\H_{J_{\eta}}$ and $R_{J_{\eta}})$ of $\Acan$. We then ask if the dynamical automorphism $\h{\Lambda}_{\eta_{2},\eta_{1}}$ on $\Acan$ can be made inner on the resulting extended state space $\Hext$. We found that if the family $J_{\eta}$ is induced by a complex structure $\J$ on $\pscov$, the answer is in the affirmative. In this precise sense, the apparent tension between the covariant and canonical frameworks is resolved.

\subsubsection{The subtlety}
\label{s3.2.2}

While the situation in the quantum theory parallels that in the classical theory in FLRW space-times, there is also an important difference. Leaves $\psext^{\eta}$ of the classical phase space are all naturally isomorphic at the \emph{kinematic} level: $\psext = \pscan \times \mathbb{R}$. This isomorphism is distinct from that provided by the symplectic flow  which makes the \emph{dynamical} automorphism $\Lambda_{\eta_{2},\eta_{1}}$ inner. In the quantum theory, on the other hand, because the complex structures $J_{\eta}$ are all distinct and $J_{\eta_{1}} - J_{\eta_{2}}$ fails to be Hilbert-Schmidt for generic FLRW metrics, there is no \emph{kinematical} identification between the Hilbert spaces $\H_{\eta}$. We only have the dynamical maps $U_{\eta_{2},\eta_{1}}$ that provide a correspondence between quantum states at different times.  That is, $\Hext$ does not have the simple product structure $\Hext = \H \times \mathbb{R}$, which would have enabled one to \emph{kinematically} identify states at different times. Rather, the total state space $\Hext$ is now a genuine bundle, where the base space is the real line $\mathbb{R}$ corresponding to time $\eta$ and the fibers are the Fock spaces $\H_{J_{\eta}}$ which (being separable Hilbert spaces) are all isomorphic but \emph{not} naturally isomorphic at the kinematic level. We have the maps
\be  \Hext \longrightarrow \mathbb{R}, \quad {\rm and}, \qquad \eta \in \mathbb{R}   \longrightarrow  \H_{J_{\eta}} \subset \Hext \, ;\ee
but kinematics does not endow the total bundle $\Hext$ with natural horizontal sub-spaces. As we discuss below, it is important to keep this distinction in mind while interpreting dynamics on $\Hext$.

From this discussion, it may appear that the contrast is tied with the transition from classical to quantum descriptions. This is not correct: the subtlety is specific to quantum field theory on \emph{time-dependent background space-times}%
\footnote{More precisely, the subtlety arises if there is no time-like Killing field whose flow preserves the chosen foliation by Cauchy surfaces. If $(\mathbb{M}, g_{ab})$ does not admit a time-like isometry, then this subtlety arises for any choice of foliation. But it also arises in stationary ---even flat--- space-times if the foliation used to describe dynamics is not preserved by any time-like Killing field \cite{torre-varadarajan}.}
and not shared by other quantum systems which also have time dependent Hamiltonians but live in a time independent space-times, or have a finite number of degrees of freedom. Consider for example non-relativistic systems with time dependent Hamiltonians. Then, in the classical description, to make the dynamical automorphism on the algebra of observables \emph{inner}, we again have to work with the extended phase space $\psext$. In the quantum theory, states are now represented by wave functions $\Psi(\vec{x}, t) \in \Hext$ which are square-integrable in $\vec{x}$ for each $t$. Dynamics is unitarily implemented in the same sense as in our discussion of quantum fields on FLRW space-times:
\be U_{t_{2},t_{1}}: \H_{t_{1}} \longrightarrow \H_{t_{2}}, \qquad 
U_{t_{2},t_{1}} \Psi_{1}(\vec{x}, t_{1})  = \Psi_{2}(\vec{x}, t_{2}) \ee
where $U_{t_{2},t_{1}}$ is the time-ordered exponential, $U_{t_{2},t_{1}}  = \mathbf{T}\,\exp \big[(1/i\hbar)\int_{t_{1}}^{t_{2}} H(t) dt \big]$. 
\footnote{ After this paper was posted on the arXiv, we learned of an elegant formulation of quantum dynamics of non-relativistic particles with time-dependent Hamiltonians, in the language of bundles discussed here \cite{asorey}. It provides new geometrical insights, especially in the case when the particle is interacting with a time dependent gauge field.} 
But now the leaves $\H_{t}$ of $\Hext$ are all naturally isomorphic \emph{kinematically}: $\H_{t} = L^{2}(\mathbb{R}, {\rm d}^{3}x)$ for all $t$. Therefore, the total state space $\Hext$ has a direct product structure just as that of the total phase space $\psext$:
\be \Hext = \H \times \mathbb{R}, \qquad {\rm with}\quad \H = L^{2}(\mathbb{R}, {\rm d}^{3}x)\, . \ee

The situation is completely analogous if we consider linear, massive quantum fields with external, \emph{time dependent} potentials, say of compact support on flat space-time with topology $\mathbb{T}^{3} \times \mathbb{R}$. Then, in the classical theory, to make the dynamical automorphism \emph{inner}, we are again led to construct $\psext$. But now we can adopt the following strategy in the quantum theory. We can just fix a complex structure $J$ on $\pscan$ (e.g., $J = J_{\rm in}$), construct the corresponding representation $(\H_{J}, R_{J})$ of $\Acan$, set the extended state space to be just
\be \Hext = \H_{J} \times \mathbb{R}  \ee
and make the dynamical automorphism $\hat{\Lambda}_{\eta_{2},\eta_{1}}$ \emph{inner} via a family of unitary maps $U_{\eta_{2},\eta_{1}}$ from states $|\Psi_{1},t_{1}\rangle$ to  $|\Psi_{2},t_{2}\rangle$. This is possible because $J- E_{\eta_{2},\eta_{1}} J E^{-1}_{\eta_{2},\eta_{1}}$ is now Hilbert-Schmidt. (This follows from the arguments used in Lemma 1 and 2 because now the volume of the 3-torus ---or, the scale-factor $a$--- is time independent.) Because the total state space $\Hext$ can be taken to be a trivial bundle, we now have a \emph{kinematical} identification between states at different times, in addition to the dynamical map $U_{\eta_{2},\eta_{1}}$ which evolves states on the leaf $\H_{J}^{\eta_{1}}$ to those on the leaf $\H_{J}^{\eta_{2}}$. Thanks to the kinematical isomorphism, it is meaningful to say that the state $|\Psi_{2},\eta_{2}\rangle = U_{\eta_{2},\eta_{1}}\, |\Psi_{1}, \eta_{1}\rangle$ at time $\eta_{2}$ is different from the state $|\Psi_{1},\eta_{1}\rangle$ at time $\eta_{1}$.

On generic FLRW backgrounds, on the other hand, because the geometry is \emph{time-dependent}, the complex structure $J_{\eta_{1}}$ is  \emph{inequivalent} to $J_{\eta_{2}}$, whence the state space is a non-trivial bundle: there is no kinematical identification between states at different times. As a result, now we only have the dynamical relation between the states at two different times given by $U_{\eta_{2},\eta_{1}}$. In the case when $J_{\eta}$ are determined by a complex structure $\J$ on $\pscov$ through $J_{\eta} = I_{\eta}\, \J \, I^{-1}_{\eta}$, the operator $U_{\eta_{2},\eta_{1}}$ maps the vacuum state in $\H_{J_{\eta_{1}}}$ to that in $\H_{J_{\eta_{2}}}$. However, 
since there is no kinematical identification between  $\H_{J_{\eta_{1}}}$ and $\H_{J_{\eta_{2}}}$, this does \emph{not} imply that the dynamics is `trivial'. The non-trivial time evolution of states in the Schr\"odinger picture can be made more explicit by writing the evolution of the two-point function of field operator (-valued-distributions). In the case under consideration in which $J_1$ and $J_2$ are induced by a single covariant complex structure $\mathfrak{J}$, the representation of the field operators can be made explicit using a 
positive frequency basis of solutions corresponding to $\mathfrak{J}$:
on $\mathcal{H}_{J_{\eta_1}}$ we have $R_{J_{1}}(\hat \varphi(\vec{x}))=\sum_{\vk} (\hat A_{\vec{k}}\, e_k(\eta_1)+\hat  A^{\dagger}_{-\vec{k}}\, e^{\star}_k(\eta_1) )\, e^{i \vec{k} \cdot \vec{x}}$, and on $\mathcal{H}_{J_{\eta_2}}$, we have $R_{J_{2}}(\hat \varphi(\vec{x})) = \sum_{\vk} (\hat A_{\vec{k}}\, e_k(\eta_2)+\hat  A^{\dagger}_{-\vec{k}}\, e^{\star}_k(\eta_2) )\, e^{i \vec{k} \cdot \vec{x}}$. Then, at time $\eta_1$, the 2-point function is given by
\be\langle 0_{1}| R_{J_{1}}(\hat \varphi(\vec{x})\, \hat \varphi(\vec{x}'))|0_1\rangle =\sum_{\vk} |e_k(\eta_1)|^2\, e^{i \vec{k} \cdot (\vec{x}-\vec{x}')}\, .\ee
Evolution in the Schr\"odinger picture  maps it to
\ba \langle U_{\eta_1,\eta_2}\, 0_{1}| R_{J_{2}}(\hat \varphi(\vec{x})\, \hat \varphi(\vec{x}'))|U_{\eta_1,\eta_2}\, 0_1\rangle &=& \langle 0_{2}|R_{J_2}( \hat \varphi(\vec{x})\, \hat \varphi(\vec{x}'))|0_2\rangle \nonumber\\
&\equiv& \sum_{\vk} |e_k(\eta_2)|^2\, e^{i \vec{k} \cdot (\vec{x}-\vec{x}')}\, .\ea
This evolution is non-trivial, since $e_k(\eta_1)$ are different from $e_k(\eta_2)$, and it agrees with the familiar evolution for the two-point function in the textbook covariant approach.\\

In this section we were primarily interested in resolving the apparent tension between the covariant and canonical descriptions. But we can also look at the problem purely from a canonical perspective. Then, to pass to the quantum theory starting from the extended phase space $\psext$ we need to equip each leaf $\psext^{\eta}$ with a suitable complex structure $J_{\eta}$. Any given (adiabatically regular) complex structure $\J$ on $\pscov$ provides a suitable family $J_{\eta}$ on $\pscan$.  But we can also choose another family $J_{\eta}$, \emph{without any reference to a complex structure $\J$ on $\pscov$}, where $J_{\eta_{2}}$ does \emph{not} equal $E_{\eta_{2},\eta_{1}} J_{\eta_{1}}E^{-1}_{\eta_{2},\eta_{1}}$. So long as $(J_{\eta_{2}} - E_{\eta_{2},\eta_{1}} J_{\eta_{1}}E^{-1}_{\eta_{2},\eta_{1}})$ is Hilbert-Schmidt for all $\eta_{1}, \eta_{2}$, we will again be led to a natural unitary implementation of the dynamical automorphisms $\h\Lambda_{\eta_{2},\eta_{1}}$. Now, the operators $U_{\eta_1, \eta_2}$ will generically map the vacuum in $\H_{J_{\eta_1}}$ to an excited state in $\H_{J_{\eta_2}}$. An example of this situation is provided by the S-matrix description discussed in section \ref{s4}. \\

\emph{Remark:} In the classical theory, the total state space $\psext = \ps\times \mathbb{R}$ is not a symplectic space. Rather, each of its leaves $\psext^{\eta}$ is a symplectic space. Similarly in the quantum theory the total state space $\Hext$ does not have a Hilbert space structure. Rather, each of its fibers, $\H_{J_{\eta}}$, has a natural Hilbert space structure. These features are common to all three cases discussed above: non-relativistic quantum mechanics with time dependent potentials, quantum field theory in flat space-time with time dependent external potentials of compact support, and quantum field theory in FLRW space-times discussed in this paper.

\subsection{Trading time dependence from space-time geometry to a potential} 
\label{s3.3}

In this sub-section, we will discuss an interesting example, of direct interest to the cosmological perturbation theory, in which a field redefinition lets us regard a test quantum field in FLRW space-time as propagating in flat space-time, but interacting with a time dependent potential. The example creates an apparent puzzle because, while in the original version the dynamical automorphism can be unitarily implemented \emph{only in the generalized sense} of section \ref{s3.2}, in the second version, dynamics can be unitarily implemented in the standard sense, i.e., within a \emph{single} representation of the canonical algebra. This issue is explained and resolved in the first part of this subsection. In the second part, we use the example to bring out the interplay between space-time geometry and the ultraviolet properties of test quantum fields in that space-time. 

\subsubsection{The issue of unitarity}
\label{s3.3.1}

Consider a massless%
\footnote{The results of this sub-section hold also in the massive case. We focus on the massless case because of its direct relevance to the cosmological perturbation theory.} 
scalar field $\phi$ on a FLRW space-time $(\mathbb{M}, g_{ab})$ where, as before, the metric is given by
\be \label{FLRW} g_{ab} \dd x^{a} \dd x^{b}\, \equiv\, a^{2}(\eta)\, \go_{ab}\dd x^{a} \dd x^{b}\, =\, a^{2}(\eta)\, (-\dd \eta^{2} + \dd {\vec x}^{2})\, , \ee
where $\go_{ab}$ is a flat metric on $\mathbb{T}^{3}\times\mathbb{R}$.  As discussed in the last two sub-sections, a complex structure $\J$ on $\pscov$ naturally induces a 1-parameter family of complex structures $J_{\eta}$ on $\pscan$, any two of which are (Hilbert-Schmidt) \emph{inequivalent}. As a result, the dynamical automorphism $\h{\Lambda}_{\eta_{2},\eta_{1}}$ fails to be unitarily implemented in any one representation $R_{J_{\eta}}$ of the canonical observable algebra $\Acan$. 

However, because FLRW space-times are conformally flat, one can adopt 
another strategy, as is often done in the cosmological perturbation theory. Recall first that because of the conformal transformation properties of the $\Box$-operator and the scalar curvature $\mathcal{R}$ of $g_{ab}$, one has
\be \Box\, \phi - \frac{\mathcal{R}}{6}\,\phi = a^{-3}\,\,{\Boxo}\, \phio \ee
because $\mathring{\mathcal{R}}$ vanishes, where we have set $\phio = a\phi$. (The the cosmological literature, $\phio$ is generally denoted by $\chi$.)  Using the fact that $\mathcal{R}$ is given by $\mathcal{R} = 6\,(a^{\prime\prime}/a^{3})$, where a prime denotes derivative with respect to $\eta$, it follows that $\phi$ satisfies $\Box\phi=0$ whenever $\phio$ is a solution of the equation 
\be \label{phioeq} \Boxo\,\phio + \frac{a^{\prime\prime}}{a}\,\phio =0. \ee
Therefore, although the physical field of interest $\phi$ propagates on the FLRW space-time, mathematically we can reduce the problem to that of a field $\phio$  propagating on a \emph{flat} space-time, $(\mathbb{M}, \go_{ab})$, in presence of a time dependent potential $(a^{\prime\prime}/a)$. 

Let us explore the quantum field theory of $\h\phio$. By repeating the constructions introduced in section \ref{s2}, one arrives at the covariant and canonical descriptions for this  field. However, because this field  `lives' in flat space-time, as discussed in  section \ref{s3.2}, a key difference arises. The 1-parameter family of complex structures $\Jo_{\eta}$ induced on $\pscano$ by any one complex structure $\calJo$ on $\pscovo$ are now (Hilbert-Schmidt) \emph{equivalent}. Therefore, there is now a natural kinematical identification between the Hilbert spaces $\H_{\Jo_{\eta}}$; the bundle $\Hext$ has a product structure $\Hext = \H_{\Jo}\, \times \mathbb{R}$, where $\Jo$ is any one of the complex structures in the family $\Jo_{\eta}$. A more interesting consequence is that the dynamical automorphisms $\Lambdao_{\eta_{2},\eta_{1}}$ on $\Acano$ are now unitarily implementable in the standard sense.%
\footnote{This follows from our results of section \ref{s3.3} and the fact that $a(\eta) = 1$ on $(\mathcal{M}, \go_{ab})$. For an alternative proof, see \cite{cortezetal}.}
For concreteness, let us fix a $\Jo$ and work with the representation $(R_{\Jo},\, \H_{\Jo})$. Then, we can state this result as follows: GivÄen any $\eta_{1},\,\eta_{2}$, there exists an unitary map $\Uo_{\eta_{2},\eta_{1}}$ on $\H_{\Jo}$ such that
\be \label{equiv3} R_{\Jo}\big(\h{\Lambdao}_{\eta_{2},\eta_{1}}(\O_{\gammao})\big) =  \Uo^{{-1}}_{\eta_{2},\eta_{1}}\, R_{\Jo}(\O_{\gammao})\, \Uo_{\eta_{2},\eta_{1}}\, , \ee
for all $\gammao \in \pscano$. Given that the relation between the two fields is very simple ---$\h{\phio}(\vec{x}, \eta) = a(\eta) \h{\phi}(\vec{x}, \eta)$--- at first sight (\ref{equiv3}) may seem to contradict the fact that the dynamical automorphism is \emph{not} unitarily implementable for the physical field $\h\phi$ in the standard sense but requires a \emph{generalized}  notion of unitarity, introduced in section \ref{s3.2}. However, this is not the case. In fact will now show that  (\ref{equiv3}) \emph{directly leads us to the generalized unitarity (\ref{implementation2})  for the physical field $\h\phi$ and vice-versa.}  

Since this issue is rather confusing at first, we will spell out the arguments in detail. A systematic analysis, starting from the Lagrangian description of the field $\phio(\vec{x}, \eta)$, shows that 
the map $\phio (\vec{x}, \eta) \rightarrow  \phi(\vec{x}, \eta) = (\phio(\vec{x}, \eta)/a(\eta))$ induces a natural 1-parameter family of symplectomorphisms $\S_{\eta}: \pscano \to \pscan$ between the two canonical phase spaces:
\be \S_{\eta} \big(\varphio (\vec{x}), \,\pio(\vec{x})\big) 
= (\varphi (\vec{x}),\, \pi(\vec{x})) := \big(a^{{-1}}(\eta)\varphio (\vec{x}), \, a(\eta)\pio(\vec{x})\big)\, ,\ee 
where the dependence on the parameter $\eta$ in $\S_{\eta}$ is a direct consequence of the fact that the scale factor $a(\eta)$ is $\eta$-dependent. More succinctly: for all $\gammao \in \pscano$, we have $\S_{\eta} (\gammao) = \gamma_{\eta} \in \psext^{\eta}$, the leaf of $\psext$ corresponding to time $\eta$. It follows immediately that the evolution maps $\Eo_{\eta_{2},\eta_{1}}$ and $E_{\eta_{2},\eta_{1}}$ on $\pscano$ and $\psext$ are related by
\be \label{E} E_{\eta_{2},\eta_{1}}\, =\, \S_{\eta_{2}}\, \Eo_{\eta_{2},\eta_{1}} \S^{-1}_{\eta_{1}}  \ee
Next, given any complex structure $\Jo$ on $\pscano$ (compatible with the symplectic structure thereon), we naturally obtain a 1-parameter family of complex structures $J_{\eta}$ on $\pscan$ (compatible with the symplectic structure thereon):
\be J_{\eta} = \S_{\eta}\, \Jo \, \S^{-1}_{\eta}\, .  \ee
The 1-particle Hilbert space $h_{Jo}$ (obtained from $(\pscano, \Jo)$) is therefore naturally isomorphic with the 1-particle Hilbert space 
$h_{J_{\eta}}$ (obtained from $(\pscan, J_{\eta})$). Denote the resulting isomorphism from the Fock space $\H_{\Jo}$ to the Fock space $\H_{J_{\eta}}$ by $V_{\eta}$. Then the representation maps $R_{\Jo}$ and $R_{J_{\eta}}$ are related by 
\be \label{R} R_{J_{\eta}} (\h\O_{\g_{\eta}})\, =\,  V_{\eta}\, R_{\Jo} (\O_{\gammao})\, V^{-1}_{\eta}\,   \ee
for all $\eta$ where $\gamma_{\eta} = \S_{\eta}\gammao$. Finally, Eqs (\ref{E}) -- (\ref{R}) and the definitions of the dynamical automorphisms on $\Acano$ and $\Acan$ imply:
\be \label{reln1} R_{\Jo}\big(\h{\Lambdao}_{\eta_{2},\eta_{1}}(\O_{\gammao})\big) =  V^{-1}_{\eta_{1}}\, \big( R_{J_{\eta_{1}}} (\h\Lambda_{\eta_{2},\eta_{1}} (\h\O_{\gamma_{\eta_{2}}}))\,\big)\, V_{\eta_{1}} \ee
With these preliminaries out of the way, we can re-express the $\h\phio$ unitarity condition (\ref{equiv3}) in terms of the quantum theory of $\h\phi$. For the left side of (\ref{equiv3}), we will use (\ref{reln1}). The right side can be re-expressed as
\be \label{reln2}\Uo^{-1}_{\eta_{2},\eta_{1}}\, R_{\Jo}(\O_{\gammao})\, \Uo_{\eta_{2},\eta_{1}} =   \Uo^{-1}_{\eta_{2},\eta_{1}}\, \big(V^{-1}_{\eta_{2}}R_{J_{2}}(\O_{\gamma_{\eta_{2}}}) V_{\eta_{2}}\big)\, \Uo_{\eta_{2},\eta_{1}} \ee
%
Let us set 
\be \label{Udef}   U_{\eta_{2},\eta_{1}} = V_{\eta_{2}}\, \Uo_{\eta_{2},\eta_{1}}\, V^{-1}_{\eta_{1}}\, .  \ee
Then, (\ref{reln1}) and (\ref{reln2}) imply that the unitarity result (\ref{equiv3}) for the mathematical $\hat\phio$ field propagating in the flat space-time metric $\go_{ab}$ \emph{is equivalent to}:
\be \label{equiv4} R_{J_{\eta_{1}}} \big(\h\Lambda_{\eta_{2},\eta_{1}}(\h{\O}_{\gamma_{1}}) \big) \, = \, U^{{-1}}_{\eta_{2},\eta_{1}}\, R_{J_{\eta_{2}}}\, U_{\eta_{2},\eta_{1}} \ee
for the physical $\h\phi$ field propagating in the FLRW space-time $g_{ab}$ of Eq. (\ref{FLRW}). But (\ref{equiv4}) is \emph{precisely the expression of generalized unitarity (\ref{implementation2}) introduced in section \ref{s3.2}.} 

To summarize, the the standard notion of unitarity for $\h\phio$ translates directly to the generalized notion of unitarity for the physical field $\h\phi$ and vice versa. Consequently, the standard quantum dynamics on the Hilbert space $\H_{\Jo}$, is equivalent to the generalized dynamics on $\Hext$. The underlying `mechanism' behind this equivalence is succinctly captured in the expression (\ref{Udef}) relating $\Uo_{\eta_{2},\eta_{1}}$ to $U_{\eta_{2},\eta_{1}}$. On the right hand side of this relation, $\Uo_{\eta_{2},\eta_{1}}$ is multiplied on the left by the kinematic map $V^{-1}_{\eta_{2}}$ determined by the value of the scale factor $a(\eta)$ at $\eta= \eta_{2}$ while on the right side it is multiplied by $V_{\eta_{1}}$ determined by the value of the scale factor at $\eta=\eta_{1}$.

\subsubsection{The issue of ultraviolet regularity}
\label{s3.3.2}

The discussion of section \ref{s3.3.1} shows that fields $\h\phio$ and $\h\phi$ are equivalent from dynamical considerations in a natural sense. The goal of this subsection is to bring out the fact  that the issue of ultraviolet regularity of operator products goes beyond unitarity of dynamics. To keep this discussion brief, we will work in the covariant description of both fields (although results can be readily translated to the canonical picture). 

Because the quantum field $\h\phio$ propagates in flat space-time, it is tempting to construct the quantum theory using structures that are familiar from flat space quantum field theory. Recall that, for a free massless field in $(\mathbb{M}, \go_{ab})$, the mode functions $e^{0}_{k}(\eta) = e^{-i k\eta}/\sqrt{2 k}$ provide the standard positive frequency basis. In presence of the time dependent potential $a^{\prime\prime}/a$, of course, they do not satisfy the field equation
\be \label{feq} \eo^{\prime\prime}_{k}(\eta) + \big(k^{2} - (a^{\prime\prime}/a)(\eta)\big) \eo_{k}(\eta)  = 0 \ee
implied by (\ref{phioeq}). Nonetheless, one can fix a convenient time $\eta=\eta_{0}$ and define a positive frequency basis $\eo_{k}(\eta)$ for the field $\phio$ by asking that it satisfy (\ref{feq}) and have the initial data corresponding to the flat space basis functions at time $\eta_{0}$, namely, $\eo_{k}(\eta_{o}) = e^{-ik\eta_{0}}/\sqrt{ 2k}$ and ${\eo}^{\prime}_{k} = -i\sqrt{k/2} \, e^{-ik\eta_{0}}$. Indeed, this strategy is commonly adopted in cosmology (see, e.g., \cite{cortezetal,cosmology}). Let us denote the complex structure on $\pscovo$ determined by this basis by $\calJo$. 
It leads one to a Fock representation $(R_{\calJo}, \H_{\calJo})$ of the algebra $\Acovo$ generated by  $\h\phio (\vec{x}, \eta)$.

However, the field of physical interest is $\h\phi$. In particular, for the back reaction calculations, we need the expectation values of the renormalized stress-energy tensor of $\h\phi (\vec{x}, \eta)$, and not of the rescaled field $\h\phio=a(\eta) \h\phi (\vec{x},\eta)$ which was introduced for mathematical convenience.  A natural strategy would be 
to induce a representation of $\Acov$ generated by $\h\phi$ starting from the representation $(R_{\Jo}, \H_{\Jo})$ of $\Acovo$. To this goal, let us begin by noting that there is a natural isomorphism $\mathcal{I}$ between the covariant phase spaces $\pscovo$ and $\pscov$ of $\phio$ and $\phi$:
\be \mathcal{I} (\phio(\vec{x}, \eta)) \, = \, \phi(\vec{x}, \eta) := 
  \phio(\vec{x}, \eta)/a(\eta)\, .  \ee
Note that, in spite of the $\eta$-dependence of the scale factor, we have a single isomorphism ---rather than a 1-parameter family of them--- because both $\phio$ and $\phi$ are functions of $\eta$. Clearly $\phi(\vec{x}, \eta)$, so defined, satisfies the desired field equation $\Box \phi =0$ and it is easy to verify that $\mathcal{I}$ preserves the symplectic product. Therefore, we can now use the complex structure $\calJo$ on $\pscovo$ to introduce a specific complex structure $\J$ on $\ps$:\,\, $J := \mathcal{I}\, \calJo\, \mathcal{I}^{-1}$. In terms of basis functions, a positive frequency basis $e_{k}(\eta)$ corresponding to $\J$ can be directly specified as solutions of 
\be e^{\prime\prime}_{k} (\eta) +  (2a^{\prime}/a)(\eta)\,\, e_{k}(\eta)  + k^{2}\,e_{k}(\eta) = 0,\ee
with initial data at $\eta=\eta_{0}$ given by
\be 
e_{k}(\eta_{0}) =\f{e^{-i k\eta_{0}}}{a(\eta_{0}) \sqrt{2k} }, \quad {\rm and} \quad e^{\prime}_{k}(\eta_{0}) = -i \sqrt{\f{k}{2}}\f{e^{-ik\eta_{0}}}{a(\eta_{0})} - \f{e^{-ik\eta_{0}}a^{\prime}(\eta_{0})}{\sqrt{2k} a^{2}(\eta_{0})}\, \ee 
(so that $e_{k}(\eta) = \eo_{k}(\eta)/a(\eta)$). A natural question is whether this basis $e_{k}(\eta)$ is regular to fourth adiabatic order so that the renormalized stress energy tensor of the $\h\phi$ field is well defined in the representation $(R_{\J}, \H_{\J})$ of $\Acov$. This can be readily checked by comparing the large $k$ behavior of these basis functions with the one needed for adiabatic regularity, given, e.g., in \cite{aan3}. 

Unfortunately, the basis fails to be regular even at second order! Thus, if we were to induce a Fock representation of $\h\phi$ from any of the seemingly natural bases $\eo_{k}(\eta)$ for $\phio$, one would not be able to renormalize dimension 2 operators such as $\h\phi^{2}(\vec{x}, \eta)$ in that representation (let alone dimension 4 operators such as the stress energy tensor). This result brings out the more general fact that considerations of ultraviolet regularity impose much more stringent requirements on the quantization procedure than the unitarity considerations. 

\section{The S-matrix Description}
\label{s4}
In this section we will show that in FLRW space-times which become time independent in the distant future and past, the generalized dynamics in the canonical theory between the {\it in} and {\it out} representations is unitary and reproduces the physics of the well-known S-matrix of the covariant theory.

Let us again consider scale factors $a(\eta)$ as in Parker's example discussed in section \ref{s2.3}, where the scale factor is constant in the past of the surface $\eta=\eta_{-}$ and to the future of another surface $\eta = \eta_{+}$ (although our S-matrix considerations will go through in a more general context, where $a(\eta)$ becomes constant only asymptotically in the distant past and the distant future at a suitable rate). In this case, in the covariant picture, one has an S-matrix description that involves \emph{two distinct} natural complex structures $\J^{\mp}$ on $\pscov$, inherited from the flat space-time metrics in the distant past and the distant future which provide the {\it in} and {\it out} Fock representations for the S-matrix. Since the scale factor $a(\eta)$ is time-independent outside a finite interval $(\eta_{1}, \eta_{2})$ (with $\eta_{1} <\eta_{-}$ and $\eta_{2} >\eta_{+}$), it follows that $(\J^{+} - \J^{-})$ is Hilbert-Schmidt (on the 1-particle Hilbert spaces $h_{\J^{-}}$ or $h_{\J^{+}}$). This immediately implies that there is a well-defined S-matrix, providing a unitary map from the past Fock space $\H_{\J^{-}}$ to the future Fock $\H_{\J^{+}}$ (see, e.g., \cite{wald-book}). As is well-known, this map  contains all the information about dynamics, in particular about particle creation and scattering amplitudes between early and late times. This situation is somewhat different from that considered in section \ref{s3} where $\pscov$ was equipped with a \emph{single} complex structure $\J$. Nonetheless, as we will now show, this S-matrix description fits-in naturally with our extended notion of unitary implementation of dynamics.

The existence of past and future flat regions provide two natural  complex structures $J_1$ and $J_2$ also on $\pscan$, directly induced by the covariant complex structures $\J^+$ and $\J-$:
\be \label{J1J2} J_{1} = I_{\eta_{1}}\, \J^{-}\, I^{-1}_{\eta_{1}}, \qquad {\rm and} \qquad  J_{2} = I_{\eta_{2}}\, \J^{+}\, I^{-1}_{\eta_{2}}\, , \ee
where $I_{\eta_1}$ and $I_{\eta_2}$ are the past and future isomorphisms between $\pscov$ and $\pscan$. Each of these complex structures on $\pscan$ provides a representation ($R_{J_{1}}$ and $R_{J_{2}}$, respectively) of the canonical algebra $\Acan$. The question is if the dynamical automorphism $\hat{\Lambda}_{\eta_{2},\eta_{1}}$ on $\Acan$ induced by the classical $S$-matrix $E_{\eta_{2},\eta_{1}}$ is unitarily implementable. More precisely, does there exist a unitary map $U_{\eta_{2},\eta_{1}}$ from the Fock space $\H_{J_{1}}$ to $\H_{J_{2}}$ such that:
\be \label{implementation3}  R_{J_{1}}\,(\hat{\Lambda}_{\eta_{2},\eta_{1}}\,\h{\O_{\gamma}}) \,  \,=\, U^{-1}_{\eta_{2},\eta_{1}}\, (R_{J_{2}} \h{\O_{\gamma}})\, U_{\eta_{2},\eta_{1}} \ee
for all $\h{\O}_{\gamma}$ in ${\Acan}$? \emph{If so, $U_{\eta_{2},\eta_{1}}$ would be the unitary S-matrix in the canonical theory.} 

But we already examined this mathematical question in a general context in section \ref{s3.2}. Our answer was that this is the case if and only if $[E_{\eta_{2},\eta_{1}}\, J_{1}\, E^{-1}_{\eta_{2},\eta_{1}} - J_{2}]$ is Hilbert-Schmidt on the 1-particle Hilbert space $h_{J_{1}}$ (see (\ref{equiv})). Let us recast this condition in terms of $\J^{-}$ and $\J^{+}$. Using (\ref{J1J2}) and the fact that the evolution map $E_{\eta_{2},\eta_{1}}$ is given by $E_{\eta_{2},\eta_{1}} = I_{\eta_{2}}\, I^{-1}_{\eta_{1}}$, we have:
\be \label{relation2} E_{\eta_{2},\eta_{1}}\, J_{1}\, E^{-1}_{\eta_{2},\eta_{1}} - J_{2} = (I_{\eta_{2}}\, I^{-1}_{\eta_{1}}) \, (I_{\eta_{1}} \J^{-} I^{-1}_{\eta_{1}})\, (I_{\eta_{1}}\, I^{-1}_{\eta_{2}})\, - I_{\eta_{2}}\, \J^{+}\, I^{-1}_{\eta_{2}} = I_{\eta_{2}}\, (\J^{-} - \J^{+})\, I^{-1}_{\eta_{2}}\, . \ee
Now, since $I_{\eta_{1}}$ is a symplectomorphism from $\pscov$ to $\pscan$, from the relation (\ref{J1J2}) between $J_{1}$ and $\J^{-}$ it follows that it extends to an isomorphism between the 1-particle Hilbert spaces $h_{\J^{-}}$ and $h_{J_{1}}$. Therefore,  (\ref{relation2}) implies that $E_{\eta_{2},\eta_{1}}\, J_{1}\, E^{-1}_{\eta_{2},\eta_{1}} - J_{2}$ is Hilbert-Schmidt on $h_{J_{2}}$ if and only if $(\J^{-} - \J^{+})$ is Hilbert-Schmidt on $\H_{J^{-}}$. But we already know this to be the case; indeed, this is the reason why the S-matrix is well-defined in the covariant theory. Thus the fact that the S-matrix is unitary in the covariant theory directly implies that $U_{\eta_{2},\eta_{1}}$, the S-matrix of the (extended) canonical theory, exists and is unitary.

\emph{Remark:} Note that, as in the extension of canonical picture developed in section \ref{s3}, the criterion for the existence of a unitary S-matrix is \emph{not} whether $(J_{2} - J_{1})$ is Hilbert-Schmidt as one might first think, but rather, whether $(J_{2} - E_{\eta_{2},\eta_{1}}\, J_{1}\, E^{-1}_{\eta_{2},\eta_{1}})$ is Hilbert-Schmidt. The meaning of this condition is rather simple: One has to apply $J_{1}$ and $J_{2}$ to the initial data at times $\eta_{1}$ and $\eta_{2}$ \emph{of the same solution} $\phi(\vec{x}, \eta)$ (rather than to the same $(\varphi(\vec{x}), \pi(\vec{x}))$ in $\pscan$).

\section{General globally hyperbolic space-times}
\label{s5}

The goal of this section is to point out that the results of previous sections admit a straightforward extension to arbitrary globally hyperbolic space-times. Specifically, while in generic situations the dynamical evolution of the  field operator and its conjugate momentum  is not unitary in any fixed Fock representation, we will show that unitarity does hold in the generalized sense of \ref{s3.2}.

Consider then a scalar field $\phi$ satisfying $\Box\,\phi - m^{2}\,\phi =0$ on a globally hyperbolic space-time $(\mathbb{M} = M, g_{ab})$. To construct the Hamiltonian description in the canonical framework, one has to introduce a time function $t$, defining a foliation of $\mathbb{M}$ by Cauchy surfaces $M_{t}$ and an evolution vector field $t^{a}$ satisfying, $t^a\nabla_a t=1$, whose integral curves identify points on  different leaves $M_{t}$. In what follows, we will work with a fixed pair $(t, t^{a})$. The covariant phase space $\pscov$ again consists of the space of solutions $\phi(\vx ,t)$ to the field equation, and the canonical phase space $\pscan$ is the space of pairs $(\varphi(\vec{x}),\, \pi(\vec{x}))$ on the 3-manifold $M$ of integral curves of $t^{a}$. As before, we can define an isomorphisms $I_{t}$  for each value of $t$, between $\pscov$ and $\pscan$:
 \ba \label{iso} I_{t}\,\phi(\vx,t)  &=& (\vp(\vx), \pi(\vx))\, \in \pscan, \qquad {\hbox{\rm where, now}} \nonumber\\
\vp(\vx) = \phi(\vx, t), &\,\,& \pi(\vx)
= \sqrt{q}\, n^a \nabla_a \phi(\vec{x},t)\, .\ea
Here $n^a$ is the future-directed, time-like  unit vector field orthogonal to $M_t$,  $q_{ab}=g_{ab}+n_an_b$ is the positive definite  metric tensor on $M_t$ induced by $g_{ab}$, and $q$ its determinant.  These isomorphisms  $I_t$ naturally define the 2-parameter family of dynamical maps $E_{t_2,t_1}$ on $\pscan$:\,\,  $E_{t_2,t_1}=I_{t_2}I^{-1}_{t_1}$. The dynamical maps, in turn define a 2-parameter family of automorphisms $\Lambda_{t_{2}, t_{1}}$ on the algebra of observables on the canonical phase space $\pscan$.

We can now construct the quantum theory following the steps laid out in subsection \ref{s2.2}. Let us begin with the abstract algebra $\Acan$ of quantum operators, generated by the (smeared versions of the) canonically conjugate pairs $(\h{\varphi}(\vec{x}),\, \h\pi(\vec{x}))$ on $M$. The classical dynamical automorphisms $\Lambda_{t_{2},t_{1}}$ naturally induce a 2-parameter family of dynamical automorphisms $\h\Lambda_{t_{2},t_{1}}$ on $\Acan$. The question now is if these are unitarily implemented in the quantum theory.

To answer this question, we need to represent the abstractly defined operators  ---smeared versions of $\h{\phi}(\vec{x}, t)$ in the covariant theory, and of $(\h\varphi(\vec{x}), \h\pi(\vec{x}))$ in the canonical theory--- by concrete operators on Hilbert spaces. In the FLRW models we constructed the necessary representations using adiabatically regular complex structures. However, since the notion of adiabatic regularity is tied to spatially homogeneous space-times, we need a more general strategy. We will replace it by the \emph{Hadamard regularity} which again leads to a natural procedure to renormalize products of operator valued distributions, such as the stress-energy tensor. Let us begin with the covariant phase space. Let $\J$, then, be a Hadamard complex structure on $\pscov$, i.e., one that leads to a Fock representation of $\Acov$ in which the $n$-point functions have the Hadamard behavior. Then, the isomorphisms $I_{\eta}$ of (\ref{iso}) again provide us with a 1-parameter family of complex structures $J_{\eta}$ on $\pscan$,\,\, $J_{\eta}=I_{\eta}\,\J \,I^{-1}_{\eta}$,\,\, which are automatically compatible with the canonical simplectic structure. These are the regular complex structures we want to consider on $\pscan$.

Now, given any complex structure $J$  in this family, the analysis of section \ref{s2.2} tells us that dynamics is unitarily implementable in the Fock space $\mathcal{H}_J$ if and only if the operator $(E_{t_2,t_1}\,J\,E_{t_2,t_1}^{-1} - J)$ is Hilbert-Schmidt on the 1-particle Hilbert space $h_{J}$. It is obvious that this condition fails on generic space-times since it  already fails for the FLRW backgrounds.  

By replacing the foliation $M_{\eta}$ used in FLRW space-times with the fixed foliation $M_{t}$ we now have, we can directly adopt the strategy presented in section \ref{s3.2} to  define a generalized notion of dynamics which {\em is} unitary. Let $\psext = \pscan \times \mathbb{R}$ denote, as before, the extended phase space, where $\mathbb{R}$ represents time. The complex structure $J_{t}$ is naturally associated with the leaf $\psext^{t}$ of $\psext$. Each $J_{t}$ again provides us with a Fock representation $(R_{J_{t}}, \H_{J_{t}})$ of $\Acan$ and we are led to an extended  space $\Hext$ of quantum states just as in the classical theory. $\Hext$ is a bundle as in section \ref{s3.2}, with the real line $\mathbb{R}$ representing time as the base space and the Fock spaces $\H_{J_{t}}$ serving as fibers. It provides us the natural arena to phrase the question of unitary implementation of the dynamical automorphism. In the extended framework, we are led to ask: Does there exist a 2-parameter family of unitary maps, $U_{t_{2},t_{1}}: \H_{J_{t_{1}}} \to \H_{J_{t_{2}}}$, that implement the dynamical automorphisms $\Lambda_{t_{2},t_{1}}$ on $\Acan$? The analysis presented in section \ref{s3.2} provides the answer:  if the family $J_t$ is induced by a single covariant complex structure $\J$ on $\pscov$ as specified above, 
then $E_{t_2,t_1}J_{t_{1}} E_{t_2,t_1}^{-1} - J_{t_{2}}$ is Hilbert-Schmidt and the required family of unitary operators $U_{t_2,t_1}$ exist.

Because the family $J_{t}$ on $\pscan$ is obtained by a single complex structure $\J$ on $\pscov$, the unitary operator $U_{t_2,t_1}$ has the property that it sends the vacuum state in $\H_{J_{t_{1}}}$ to that 
in $\H_{J_{t_{2}}}$:\,\, $U_{t_2, t_1}|0_{t_{1}}\rangle = |0_{t_{2}}\rangle$. However, this does not mean that dynamics is trivial because $|0_{t_{1}}\rangle$ and $|0_{t_{2}}\rangle$ are distinct positive linear (or expectation value) functionals on the algebra $\Acan$. In particular, by repeating the argument presented in section \ref{s3.3}, one can show that $U_{t_{2},t_{1}}$ correctly evolves the 2-point functions. More generally, if the complex structures $J_{t}$ are selected by other physical considerations and do not descend from a single $\J$ on $\pscov$, then $U_{t_2t_1}|0_{t_{1}}\rangle \not= |0_{t_{2}}\rangle$. An interesting example is provided by a metric $g_{ab}$ which becomes asymptotically stationary in the distant past or distant future as in section V. In that case, the covariant phase space $\pscov$ is naturally equipped with distinct two complex structures $\J^{\pm}$ and they induce two complex structures $J^{\pm}$ on $\pscan$. If $M_{t_{1}}$ and $M_{t_{2}}$ lie in the stationary regions in the asymptotic past and future, respectively, the map $U_{t_{2},t_{1}}$ constructed above provides us with a non-trivial S-matrix in the canonical theory.

To summarize, the main results presented in sections \ref{s3.1}, \ref{s3.2} and \ref{s4} for FLRW space-times naturally extend to quantum fields on globally hyperbolic space-times for any given choice of the foliation $M_{t}$ by Cauchy surfaces and a dynamical vector field $t^{a}$ providing an identification between them. \\

\textit{Remark:}  Already in Minkowski space, one can introduce a foliation that is \emph{not} preserved by any time-like Killing field. An early insightful analysis due to Torre and Varadarajan \cite{torre-varadarajan} showed that the corresponding dynamical automorphisms fail to be unitarily implemented in the standard Fock representation of linear fields (if the space-time dimension is greater than 2); thus the Schr\"odinger picture does not exist for the `bubble-time evolution'. However, it follows from our discussion that these automorphisms are in fact unitarily implemented on the \emph{extended} state space state $\Hext$ induced by the standard Fock representation of $\Acov$. Thus the Schr\"odinger representation does exist in the generalized sense and correctly captures the dynamical evolution of fields.

\section{Discussion}
\label{s6}
 
Recall that the textbook quantization of linear fields $\phi$ on a globally hyperbolic space-time requires an external input: a positive and negative frequency decomposition of classical fields that enables one to decompose the field operator $\h\phi$ into creation and annihilation operators as in (\ref{fo}). More succinctly, the necessary input is a complex structure $\J$ on the space of classical solutions, i.e.\ the covariant phase space $\pscov$, which is compatible with the symplectic structure thereon. Now, if the space-time is static, the time-translation isometry flow in space-time provides a natural, dynamical automorphism $\h\Lambda_{\eta_{2},\eta_{1}}$ on the canonical algebra $\Acan$ generated by the pairs $(\h\varphi(\vec{x}), \h\pi(\vec{x}))$. It turns out that one can single out the required complex structure $\J$ \emph{uniquely} by demanding that $\h\Lambda_{\eta_{2},\eta_{1}}$ be unitarily represented in the resulting Fock representation \cite{am1}.%
\footnote{We assume that the norm of the static Killing field is bounded below by some $\epsilon >0$. These considerations generalize to stationary space-times if the canonically conjugate operators and the dynamical automorphism refer to a foliation that is is preserved by the flow of the Killing field.}
Thus, the presence of  the time-translation isometry provides us with a natural Fock quantization of $\phi$ in which dynamics is satisfactorily represented. But physical considerations also lead to another, independent requirement on the viability of the quantum theory: the ultraviolet regularity that is necessary to renormalize products of operators such as the stress-energy tensor. In static space-times, this requirement is automatically satisfied on the Fock representation selected by the isometry. Thus we have the happy situation that there exists a unique Fock representation of the quantum field on which the two independent requirements are met.

In time dependent space-times, the situation is more subtle. Let us begin with the FLRW space-times that are widely used in cosmology. The mainstream strategy is to focus on ultraviolet regularity. Thus, one chooses \emph{any} complex structure $\J$ on $\pscov$ that has regularity of sufficient adiabatic order, so that the expectation values of products of operators of interest are well-defined on (a dense subspace of) the resulting Fock space. For concreteness, let us focus on order 4 regularity needed for renormalization of the stress-energy tensor (which is a dimension 4 operator). Then each regular complex structure $\J$ on $\pscov$ leads to a representation of the field algebra on a Fock space $\H_{\J}$. Since there is an infinite family of such representations, there is no preferred vacuum state or particle number operator. However all these representations are unitarily equivalent and, in cosmological applications, the power spectrum and expectation values of the renormalized stress-energy tensor are well defined in all of them. Therefore, mathematically the theory is deemed to be complete.

What about the unitarity of dynamics? Can one, as in the static case, narrow down the choice of Fock representations using unitarity considerations?  Because of spatial homogeneity, FLRW space-times have a preferred foliation which provides a natural notion of time evolution. Dynamics is most transparent in the canonical picture, especially because, now, the space-time geometry is itself time-dependent.  As discussed in section \ref{s2}, there is again a well-defined dynamical automorphism $\h\Lambda_{\eta_{2},\eta_{1}}$ on the algebra $\Acan$ generated by the canonically conjugate pairs $(\h\varphi(\vec{x}), \h\pi(\vec{x}))$. Thus, we are led to ask: Is there a complex structure $J$ on the canonical phase space such that $\h\Lambda_{\eta_{2},\eta_{1}}$ is unitarily implemented on the resulting Fock-space $\H_{J}$. More concretely, is there a Fock representation
of $\Acan$ admitting a family of unitary operators $U_{\eta_{2},\eta_{1}}$ on $\H_{J}$ such that
\be \label{unitary3}\R_{J}(\h\Lambda_{\eta_{2},\eta_{1}}\, (\h\O))  = U^{-1}_{\eta_{2},\eta_{1}}\, \R_{J}(\h\O) \,\, U_{\eta_{2},\eta_{1}}\,  \ee
for all $\O \in \Acan$? (Here $\R_{J}(\h\O)$ is the concrete representation on Fock space $\H_{J}$ of the element $\O \in \Acan$.) If so, $\h\Lambda_{\eta_{2},\eta_{1}}$ will be realized as an \emph{inner} automorphism in the representation $(R_{J}, \H_{J})$. As we explicitly showed in section \ref{s2.3}, surprisingly, \emph{the answer is in the negative}: the dynamical automorphism on $\Acan$ can not be made unitary for any $J$ \cite{cortezetal}. Thus, the natural seeming strategy to narrow down the choice of a Fock representation using unitarity of dynamics fails. 

To summarize, then, in the Heisenberg picture, dynamics is well-defined via automorphism $\h\Lambda_{\eta_{2},\eta_{1}}$ on the observable algebra and we can construct Fock representations which are ultraviolet regular. However, the standard notion of unitary dynamics fails \emph{in all these representations}; in none of them can we pass to the Schr\"odinger picture in which dynamics is transferred to states. Thus, a fundamental tenet of quantum field theory in flat (or, more generally, stationary) space-times is violated in quantum field theory on FLRW backgrounds.

To resolve this tension, in section \ref{s2.1} we re-examined the situation in the classical theory. There is again a well defined dynamical automorphism $\Lambda_{\eta_{2},\eta_{1}}$ on the algebra of classical observables. To make it \emph{inner}, one needs a Hamiltonian flow on the phase space ---the analog of the unitary maps in (\ref{unitary3})--- which evolves states and induces this automorphism on observables. As is well known, such a flow does not exist on $\pscan$; the situation is completely parallel to that in the quantum theory. But it is also well-known that the required flow does exist on the extended phase space $\psext = \pscan \times \mathbb{R}$. This suggests that the unitary implementation of the dynamical automorphism $\h\Lambda_{\eta_{2},\eta_{1}}$ in the quantum theory may be possible if one appropriately extends the state space.

In section \ref{s3}, we showed that this is indeed the case. Furthermore the required extension naturally descends from the standard covariant theory. A regular complex structure $\J$ on $\pscov$ does not naturally induce a complex structure $J$ on the canonical phase space. Rather, it induces a complex structure $J_{\eta}$ on each leaf $\psext^{\eta}$ of the extended phase space $\psext$. Therefore, the Fock representation $\H_{\J}$ of $\Acov$ naturally induces a 1-parameter family of representations of $\Acan$ on Hilbert spaces $\H_{J_{\eta}}$.
As in the classical theory, then, we are led to formulate the question of unitarity on an \emph{extended} state space $\Hext$: 
Do there exist unitary operators $U_{\eta_{2},\eta_{1}}: \,\, \H_{\eta_{1}} \to \H_{\eta_{2}}$ such that 
\be \label{implementation4}  R_{J_{1}}\,(\hat{\Lambda}_{\eta_{2},\eta_{1}}\,\h{\O}) \,   \,=\, U^{{-1}}_{\eta_{2},\eta_{1}}\, (R_{J_{2}} (\h{\O}))\, U_{\eta_{2},\eta_{1}}\,   \ee
for all $\O \in \Acan$? In section \ref{s3.2.1} we showed that the answer is in the affirmative. To summarize, in both classical and quantum theories the dynamical automorphisms fail to be \emph{inner} on the `traditional' state spaces $\pscan$ and $\H_{J}$. To make them \emph{inner} one simply needs to extend the state spaces to $\psext$ and $\Hext$ and formulate the question in terms of them. Once the extension is done, the tension disappears and we can use any complex structure $\J$ on $\pscov$ that is ultraviolet regular; dynamics in the canonical picture is guaranteed to be unitary in the naturally extended sense specified in (\ref{implementation4}).

However, as we discussed in section \ref{s3.2.2}, there is a subtle but important difference between the classical and quantum theories: while there is an obvious kinematical isomorphism between the leaves $\psext^{\eta}$ of $\psext$,  there is no such map between the leaves $\H_{\eta}$ of $\Hext$. The extended state space $\Hext$ is \emph{not} a product space of the type  $\Hext = \H\times \mathbb{R}$. It is a \emph{non-trivial bundle}, $\mathbb{R} \to \Hext$, whose base space is the real line coordinatized by $\eta$ (and representing the time-axis), with fibers $\H_{\eta}$. $\Hext$ does not admit natural horizontal subspaces while $\psext$ does. On $\psext$, the natural horizontal subspaces provide a kinematical identification between states at different times, so one can readily tell if the dynamical evolution is trivial or non-trivial. In quantum theory, the absence of a kinematical identification makes the non-triviality of dynamics much more subtle in the Schr\"odinger picture, as discussed below. This important feature is specific to situations in which space-time geometry is time-dependent in the sense that the chosen constant time slices are \emph{not} related by an isometry.

From a purely canonical perspective, the complex structures $J_{\eta}$ on the leaves $\psext^{\eta}$ do not have to descend from a complex structure $\J$; unitarity holds as long as the family $J_{\eta}$  is compatible in the sense of (\ref{equiv}).
But when they do descend from a single $\J$ on $\pscov$, the unitary map $U_{\eta_{2},\eta_{1}}$ has the counter-intuitive property that it maps the vacuum state in $\H_{\eta_{1}}$ to the vacuum state in $\H_{\eta_{2}}$:\,\, $U_{\eta_{2},\eta_{1}}\, |0_{\eta_{1}}\rangle = |0_{\eta_{2}}\rangle$. However, because there is no kinematical isomorphism between the two Hilbert spaces, the state $|0_{\eta_{1}}\rangle$ is \emph{not} the same as the state $|0_{\eta_{2}}\rangle$: they define \emph{distinct} positive linear (or expectation value) functionals on the algebra $\Acan$. To make this point explicit, we calculated  expectation values of the operator $\h\varphi(\vec{x})\, \h\varphi(\vec{x}^{\prime})$  in the two states and showed that this 2-point function does evolve non-trivially from time $\eta_{1}$ to time $\eta_{2}$, and the evolution is exactly what it should be, from the Heisenberg picture, which is well defined a priori.

In section \ref{s3.3}, we considered an interesting interplay in the FLRW space-times that shows that our extension of the notion of unitarity is inevitable. Because FLRW space-times are conformally flat, the field $\phi$ satisfies $\Box \phi =0$ with respect to the FLRW metric $g_{ab}$ if and only if $\phio := a(\eta)\phi$ satisfies $(\Boxo + V(\eta)) \phio = 0$ with respect to the \emph{flat metric} $\go_{ab} = a^{-2}(\eta)\, g_{ab}$, where $V(\eta) = (a^{\prime\prime}/a)$ is a time-dependent external potential. Thus, in place of a field $\phi$ living in the time dependent FLRW space-time, we can study the field $\phio$ that lives in \emph{flat space-time}, but interacts with $V(\eta)$. Now, in the flat metric $\go_{ab}$, the $\eta={\rm const}$ surfaces are related by a time-translation isometry. Therefore, a key simplification occurs: One can introduce a single complex structure $\Jo$ on $\pscano$, construct the Fock representation of $\Acano$, and show that the dynamical automorphism on $\Acano$ is made \emph{inner} by a family of unitary operators $\Uo_{\eta_{2},\eta_{1}}$ on $\H_{\Jo}$ \cite{cortezetal}. Thus, for the $\phio$ field, one does not have to construct an extended state space or extend the notion of unitarity. Because the relation $\phio(\vec{x},\eta) = a(\eta) \phi(\vec{x},\eta)$ is so simple, we can translate the result to the dynamics of the $\h\phi$ field. We showed that the translation to the $\h\phi$ field is \emph{precisely the notion of extended unitarity of (\ref{implementation4})!} Because of this complete equivalence, in the quantization procedure one can work with either fields. However, since the physical field is $\phi$, it is important to ensure that the Fock representation one constructs is sufficiently regular in the ultraviolet so that the expectation values of the stress-energy tensor operator of $\h\phi$ are well-defined. As we showed in section \ref{s3.3.2}, \emph{natural seeming quantization procedures starting with $\phio$ can fail to capture this physical requirement}. This problem can be readily alleviated by working directly with $\phi$.
 
In section \ref{s4} we discussed the S-matrix theory in the case when the scale factor becomes time independent in the distant past and the distant future. In the covariant, textbook picture it is well-known that there exist {\em in} and {\em out} Hilbert spaces $\H_{\J^{\mp}}$ of asymptotic states, determined by the past and future complex structures $\J^{\mp}$ on $\pscov$, and a well defined unitary S-matrix  which maps $\H_{\J^{-}}$ to $\H_{\J^{+}}$. This situation is different from that considered in section \ref{s3} because we now have two different complex structures on $\pscov$ ---i.e. two distinct decompositions of field operators into creation and annihilation parts--- rather than just one. Nonetheless, considerations of section \ref{s3} naturally apply also to this case and provide us with a unitary S-matrix in the extended canonical description. 

In section \ref{s5} we sketched a generalization of our constructions to globally hyperbolic space-times equipped with a foliation $M_{t}$ by Cauchy surfaces and an evolution vector field $t^{a}$. Again, there is no tension between the covariant framework and the extended canonical framework. The central lesson can be summarized as follows. In globally hyperbolic space-times, one can always introduce regular Fock representations of the algebra of observables of linear quantum fields in which dynamics is well defined in the Heisenberg picture. However, if the chosen foliation by Cauchy surfaces is not left invariant by a time-translation isometry ---as is always the case in non-stationary space-times-- then the passage to the Schr\"odinger picture is more subtle. One has to construct a 1-parameter family of representations $(R_{t}, \H_{t})$ of the canonical quantum algebra $\Acan$, one for each leaf of the $t= {\rm const}$ foliation. Typically, these representations are unitarily inequivalent. Still, the dynamical automorphisms $\h\Lambda_{t_{2},t_{1}}$ \emph{are} implemented by unitary maps $U_{t_{2},t_{1}}$ from $\H_{t_{1}}$ to $\H_{t_{2}}$ such that the evolution of expectation values obtained by evolving states in this (extended) Schr\"odinger picture is the same as that in the Heisenberg picture which was well defined from the beginning.\\

We will conclude with a remark that suggests a direction for future investigations.

In the case when the geometry is asymptotically stationary in the distant past and distant future, one can introduce two natural representations $(R_{\J^{\mp}}, H_{\J^{\pm}})$ of the covariant observable algebra $\Acov$ which then naturally descend to the representations of the canonical algebra $\Acan$. As we saw in section \ref{s4}, this makes the dynamics encoded in the S-matrix transparent. What about more general contexts where one does not have static asymptotic regions? Given a foliated globally hyperbolic space-time, can one perhaps naturally associate a representation $(R_{\J_{t}}, H_{\J_{t}})$ with each leaf $t={\rm const}$ of the foliation? In specific geometries, physical considerations could provide the required family $\J_{t}$ of complex structures on $\pscov$. Consider, for instance, an idealized but instructive example in which the space-time geometry is flat in a small time interval of length $\epsilon$, then becomes dynamical in an interval of length $\delta$ and repeats this behavior. Then for each $t$ in the $\epsilon$-interval in which the  space-time metric is flat, there would be a natural complex structure. If we restrict ourselves to points $t_{n}$, each lying in such an $\epsilon$-interval, then the dynamical automorphims $\h{\Lambda}_{t_{n},t_{m}}$ will be unitarily implementable on the corresponding $\Hext$. In the most commonly used FLRW space-times, one can generalize this procedure and associate a preferred `instantaneous' complex structure $\J_{\eta}$ on $\pscov$ \cite{ana}.%
\footnote{The associated representation $R_{J_{\eta}}$ is characterized by the property that, in the adiabatic regularization scheme, the vacuum expectation value of the renormalized stress-energy tensor vanishes at that instant $\eta$.  Our results of sections \ref{s2.3} and \ref{s3.3} imply that in FLRW space-times, the `instantaneous Hamiltonian diagonalization' procedure, that was advocated in the older literature, will also lead to generalized unitarity, although states in these representations will not be regular to second adiabatic order. In more general space-times the procedure can become ambiguous \cite{ana}.}
In this case, dynamics is again unitary in the extended canonical picture even though now $J_{\eta} \not= E_{\eta,\eta^{\prime}}\, J_{\eta^{\prime}}\, E^{{-1}}_{\eta,\eta^{\prime}}$ (because $\J_{\eta} \not= \J_{\eta^{\prime}}$). Thus, the Schr\"odinger picture exists in the extended setting and there is complete harmony between the covariant and canonical descriptions. Is there perhaps a way to extend such constructions to general space-times? While this is unnecessary from the perspective of a purely algebraic approach to quantum field theory in curved space-times, such constructions are often useful for an intuitive understanding of the physical processes that occur because of the interaction of the quantum field with the curved space-time geometry.

\section*{Acknowledgments}

We would like to thank Alejandro Corichi, Guillermo Mena Marugan, Alok Ladha and Javier Olmedo for stimulating discussions, and Guillermo Mena Marugan and Jose Vehlinho for clarifying correspondence. This work was supported by the NSF grants PHY-1403943 and PHY-1205388, the Eberly research funds of Penn State.

\end{document}